\documentclass[11pt]{article}

\usepackage[margin=1.05in]{geometry}
\usepackage{amsmath,amssymb,amsthm,mathtools}
\usepackage{enumitem}
\usepackage{microtype}
\usepackage[colorlinks=true,linkcolor=blue]{hyperref}
\usepackage{bbm}
\usepackage{graphicx}
\usepackage{booktabs}
\usepackage{standalone}
\usepackage{tikz}
\usepackage{algorithm}
\usepackage{algpseudocode}
\usetikzlibrary{arrows.meta,calc,decorations.pathreplacing,backgrounds,positioning}

\usepackage{natbib}

\hypersetup{
    colorlinks=true,
    citecolor=gray
}

\usepackage{authblk}

\usepackage{thm-restate}

\allowdisplaybreaks
\usepackage{setspace}

\theoremstyle{plain}
\newtheorem{theorem}{Theorem}[section]
\newtheorem{lemma}[theorem]{Lemma}
\newtheorem{proposition}[theorem]{Proposition}
\newtheorem{corollary}[theorem]{Corollary}
\theoremstyle{definition}
\newtheorem{definition}[theorem]{Definition}
\theoremstyle{remark}

\newcommand{\LongState}[1]{%
  \State \parbox[t]{\dimexpr\linewidth-\algorithmicindent\relax}{#1}%
  \vspace{0.2\baselineskip}%
}

\newcommand{\remove}[1]{}

\newcommand{\mpmd}{\operatorname{MPMD}}
\newcommand{\ALG}{\text{\textsc{alg}}}
\newcommand{\OPT}{\text{\textsc{opt}}}
\newcommand{\Dual}{\text{\textsc{dual}}}
\newcommand{\Delay}{\text{\textsc{delay}}}
\newcommand{\reach}{\operatorname{reach}}

\newcommand{\sumcost}{\operatorname{sum}}

\newcommand{\len}{\operatorname{len}}
\newcommand{\age}{\operatorname{age}}
\newcommand{\wait}{\operatorname{wait}}
\newcommand{\merge}[1]{\mu_{#1}}
\newcommand{\bd}{\partial}
\newcommand{\cH}{\mathcal H}

\newcommand{\cA}{\mathcal A}
\newcommand{\cC}{\mathcal C}
\newcommand{\cD}{\mathcal D}

\newcommand{\bbR}{\mathbb{R}}
\DeclarePairedDelimiter{\abs}{\lvert}{\rvert}
\newcommand{\cost}{\mathrm{cost}}

\title{A General Framework for Matching with Delays}

\author[1]{Sara Ahmadian}
\author[1,2]{Shuchi Chawla}
\author[1]{Ravi Kumar}
\author[1]{Manish Purohit}
\author[3]{Shirley Zhang}
\affil[1]{Google Research}
\affil[2]{The University of Texas at Austin}
\affil[3]{Harvard University}

\date{}

\begin{document}
\maketitle

\begin{abstract}
We present a simple deterministic $O(\log n)$-competitive algorithm for the online minimum-cost perfect matching with delays (MPMD) problem on arbitrary, unknown metric spaces, where $n$ is the (unknown) number of requests. This improves upon the recent $O(\log^5 n)$-competitive algorithm of Dufay and Wattenhofer (SODA 2026) and nearly matches the $\Omega(\log n / \log \log n)$ randomized lower bound of Ashlagi et al.\ (APPROX 2017).
Our algorithmic framework rests on two ingredients: an online moat-growing process that grows active dual variables at a deliberately slowed constant rate $\alpha < 1$ to control temporal asymmetry, and a delayed eligibility rule that matches requests in the same dual component only after their waiting time is proportional to their time-expanded distance. The framework extends naturally to arbitrary concave delay functions, maintaining an $O(\log n)$ competitive ratio, and to $k$-way matching with delays, yielding an $O(k \log n)$-competitive deterministic algorithm.
\end{abstract}

\thispagestyle{empty}
\addtocounter{page}{-1}

\newpage

\section{Introduction}
\label{sec:intro}

In this paper we study the problem of  \emph{min-cost perfect matching with delays} (MPMD), introduced by~\cite{emek2016online}. In this problem, requests arrive over time in a metric space. The algorithm may irrevocably match any two pending requests at any time, paying the metric distance between their locations together with the time each has waited since its arrival. Every request must eventually be matched, and the objective is to minimize the total connection and delay cost. 

Delay in service is an important consideration in many online systems. However, postponing service can sometimes create opportunities to improve its quality or use resources more efficiently. This tradeoff arises naturally in matching problems, where requests are processed in pairs. For example, an online gaming platform seeks to pair players of similar skill to provide a balanced and engaging game. A ride-pooling service combines requests with compatible routes to reduce travel costs and passenger detours. A computational system groups jobs with similar processing requirements into batches to reduce wasted capacity. In each setting, a suitable partner may not be available when a request arrives, and waiting for future arrivals can improve the match. But this flexibility sets up a tradeoff: matching immediately can commit the system to poor choices, while waiting too long can erase the benefit of a better match. The algorithmic challenge is to manage this tradeoff without knowing which requests will arrive next.

\paragraph{Main contributions}

We study $\mpmd$ in the general setting when neither the metric space (which may be infinite) nor the number of requests ($n$) is known a priori: the algorithm learns only the locations of arriving requests and the distances between them.  We develop a new framework that yields simple deterministic algorithms with nearly tight competitive ratios. The framework rests on two novel ideas: an online dual-growing process that grows dual variables at a deliberately reduced rate, and a delayed eligibility rule that allows requests to be matched only once they belong to the same dual component and satisfy a waiting time threshold. Together, these ideas yield a surprisingly simple algorithm: \textsl{match eligible pairs as soon as possible, breaking ties arbitrarily}.

For this unknown-metric setting,~\cite{bienkowski2017match} first gave an $O(n^{\log_2 5.5})$-competitive algorithm; this was later improved to $O(n)$ by \cite{bienkowski2018primal} using the same primal-dual formulation that we use in this paper. Subsequently, \cite{azar2020deterministic} reduced this factor to $O(n^{\log_2 1.5+\epsilon}) \approx O(n^{0.59})$. In a recent breakthrough, \cite{dufay2026deterministic} obtained the first deterministic polylogarithmic competitive ratio of $O(\log^5 n)$.  We show:
\begin{theorem}[Main result]
\label{thm:intro-main}
There is a deterministic $O(\log n)$-competitive algorithm for min-cost perfect matching with delays on arbitrary metric spaces. The algorithm requires no advance knowledge of the metric space or the number of requests.
\end{theorem}
Our bound is nearly tight, as \cite{ashlagi2017min} showed that no algorithm, even randomized, can obtain a ratio better than $\Omega(\log n/\log\log n)$ against an oblivious adversary. As we will see, relative to the earlier primal-dual approach, a key difference in our work is that the dual grows at a slow constant rate below $1$; this allows the primal--dual gap to be bounded by a constant.

\paragraph{Extensions} The same approach extends with only minor changes to two natural variants. First, suppose that a request waiting for $t$ units of time incurs cost $\phi(t)$, where $\phi$ is a known, increasing, unbounded concave function with $\phi(0)=0$.  MPMD with concave delays has also been well studied in literature and \cite{deryckere2023set} gave an $O(n)$-competitive algorithm.  Adjusting the dual growth and matching thresholds to the delay function $\phi$, we show:
\begin{theorem}[Concave delays]
\label{thm:intro-concave}
There is a deterministic $O(\log n)$-competitive algorithm for min-cost perfect matching with arbitrary concave delays on arbitrary metric spaces.
\end{theorem}
Our techniques also naturally extend to the $k$-way matching problem with delays, in which requests must be partitioned into groups of size exactly $k$.  The problem was introduced by~\cite{melnyk2021kway} to model multiplayer online games.  When the connection cost of a group is equal to the sum of all the pairwise costs inside it, and the delays are linear, we show: 
\begin{theorem}[$k$-way matching]
\label{thm:intro-kway}
There is a deterministic $O(k \log n)$-competitive algorithm for min-cost $k$-way matching with delays on arbitrary metric spaces with sum group costs.
\end{theorem}
Thus, for every fixed $k$, the dependence on the number of requests remains logarithmic. Through a reduction to sum costs, we extend our framework to several other metrics on sets of size $k$, obtaining $O(\operatorname{poly}k\cdot\log n)$ competitive ratios. These include diameter, MST cost, and the general class of $H$-metrics introduced by \cite{melnyk2021kway} to model $k$-way connection costs. See Appendix~\ref{app:kway} for these extensions.

\subsection{Key technical ideas}

Our starting point is the observation of \cite{dufay2026deterministic} that the cost of a greedy matching can be charged within a logarithmic factor to the cost of a minimum spanning tree (MST) on all requests. The problem is that this MST can be polynomially more expensive than a min-cost matching. This raises a natural question: can we partition the requests online into components so that the sum of the MST costs of the components is comparable to the minimum matching cost? They construct such a partition using a Bor\r{u}vka-style distributed greedy procedure; we instead identify the partition through an online dual-growing process.

\paragraph{A thought experiment}
Suppose we could run an offline moat-growing dual construction process of \cite{goemans1995general} in a ``time-expanded'' metric (defined in Section \ref{sec:prelims}), and bound the MST cost of each resulting component by the total dual value accumulated within it. 
Running greedy matching separately within each final component would therefore yield a logarithmic approximation.

While tempting, there are a few problems with this approach. First, the component-wise MST bound does not hold: the MST of a component can cost much more than the dual value accumulated within it. Second, since the dual is grown online, matching decisions must be made while the dual is still evolving and before the final partition into components is known. It is not clear how to implement an (approximately) greedy matching within each eventual dual component.

The third obstacle is the most fundamental: online dual growth is inherently \emph{asymmetric} in time. Consider, for example, two requests $u$ and $v$, with $v$ arriving after $u$. The dual associated with $u$ begins growing earlier and can therefore pay a disproportionately large share of the cost of the edge $(u, v)$. In the extreme case where the two requests occupy the same location in the underlying metric, the dual associated with $u$ may already pay the entire time-expanded cost of $(u, v)$ when $v$ arrives. This asymmetry can leave the online dual polynomially smaller than the optimal matching cost; we give an example in Appendix \ref{sec:fast-dual-counterexample}. 

\paragraph{Our approach}
Our central idea is to slow down the online dual growth. This controls temporal asymmetry and, together with a delayed matching rule, supports a direct charging argument that bypasses the problematic MST bound. In particular, it immediately limits the temporal imbalance in the example above. Suppose, for example, that each active dual grows at rate $1/2$. By the time $v$ arrives, the dual associated with $u$ can then pay at most half the cost of $(u, v)$, leaving room for both duals to grow simultaneously after $v$ arrives. We show that for any fixed growth rate strictly in $(0, 1)$, the resulting dual value is within a constant factor of the primal optimum (Appendix \ref{sec:fast-dual-counterexample}). 

We combine slow dual growth with a delayed matching rule adapted from \cite{azar2020deterministic,dufay2026deterministic}. Two unmatched requests become eligible to be matched once they belong to the same dual component and the time elapsed since the earlier arrival is at least twice their distance in the time-expanded metric. The algorithm matches eligible pairs immediately, breaking ties arbitrarily. This delay gives nearby requests time to arrive and allows further dual growth to bring requests in different components together.

Although our resulting matching is not exactly greedy, it retains a key structural property: sufficiently old requests that remain unmatched within the same dual component must be pairwise separated at a distance proportional to their age. 
Our main technical step is to exploit this separation property and the slow dual growth to show that enough dual value has accumulated in the recent past to pay for these requests at the corresponding scale. This yields a direct charge of the number of sufficiently old unmatched requests, weighted by their age scale, to recent dual growth, circumventing the need to bound the MST cost of the entire component. This local charging property (Theorem~\ref{thm:recent-window}) is the technical crux of our competitive ratio analysis.

Our eventual algorithm bears some similarity to the $O(n)$-competitive online primal--dual algorithm of  \cite{bienkowski2018primal}: both use moat growing to construct an online dual solution for the same LP formulation and match unmatched requests within the same dual component. However there are two crucial differences: (i) we slow dual growth to a rate of $\alpha<1$ and (ii) we defer matching until sufficient time has elapsed, rather than matching requests as soon as they enter the same component. Together, these changes yield an exponential improvement in the competitive ratio, from $O(n)$ to $O(\log n)$.

Notably, our approach does not extend directly to \emph{bipartite} matching with delays, where requests have two types and only opposite types may be paired. We leave obtaining a logarithmic guarantee for online bipartite matching as an open question.

\subsection{Related work}

In the following, let $N$ be the number of points in the underlying finite metric space.

\paragraph{Matching with linear delays}
Early work on $\mpmd$ focused on the case when the metric space is known to the algorithm. \cite{emek2016online} introduced the MPMD problem and gave a randomized $O(\log^2 N+\log\Delta)$-competitive algorithm, where $\Delta$ is the aspect ratio of the metric space; \cite{azar2017polylogarithmic} improved this to $O(\log N)$. Both of these algorithms use probabilistic tree embeddings to embed the metric space into an HST and are thus randomized. Since $N$ can be much larger than $n$, or even unbounded, these results do not imply any bounds in terms of $n$. \cite{ashlagi2017min} proved an $\Omega(\log N/\log\log N)$ lower bound for randomized algorithms. Their construction uses $\Theta(N)$ requests, yielding the lower bound in terms of $n$ stated above.

\paragraph{Bipartite matching}
The bipartite version of the problem has also been studied extensively. ~\cite{ashlagi2017min} first developed a randomized $O(\log N)$-competitive algorithm on known finite metrics and an $\Omega(\sqrt{\log N/\log\log N})$ randomized lower bound. The deterministic algorithms of ~\cite{bienkowski2018primal} and ~\cite{azar2020deterministic} also apply to this variant with their respective guarantees. A competitive ratio polylogarithmic in $n$ remains open.

\paragraph{Nonlinear and set-dependent delays}
\cite{azar2021concave} gave randomized $O(\log N)$-competitive algorithms for concave delays in both the non-bipartite and bipartite settings.  \cite{deryckere2023set} gave a deterministic $O(n)$-competitive algorithm for a common concave delay function without requiring advance knowledge of the metric; our concave-delay extension improves this dependence to $O(\log n)$. More general delay models have also been studied:~\cite{liu2018impatient} considered convex delays,~\cite{deryckere2023set} introduced delays depending on the set of pending requests, and~\cite{gan2026size} obtained further bounds for size-based and convex delays. 

\paragraph{$k$-way matching}
\cite{melnyk2021kway} introduced $k$-way matching with delays and the class of $H$-metrics%
\footnote{An \emph{$H$-metric} is any symmetric, non-negative function $H : X^k \rightarrow [0, \infty)$ that assigns a cost to $k$ requests and satisfies certain monotonicity and generalized triangle inequality constraints.}%
, obtaining a randomized $O(k^5\log N)$ bound. \cite{kakimura2025kway} gave a deterministic primal-dual algorithm with competitive ratio $O(k^2 n)$ for $H$-metrics, improving to $O(n+k^2)$ when group costs are diameters on the line. Our extension gives $O(k^2\log n)$ for diameter costs in arbitrary metric spaces and also implies an $O(k^3 \log n)$-competitive algorithm for general $H$-metrics.

\paragraph{Special cases} Stronger guarantees are known for the above variants in several special cases. \citet{emek2019minimum} give an optimal deterministic $3$-competitive algorithm for two-point metrics, while~\cite{azar2017polylogarithmic} obtain a deterministic $O(h)$-competitive algorithm on tree metrics of height $h$.  \cite{liu2018impatient} study the convex delay version over uniform metrics, obtaining an asymptotically optimal deterministic $O(N)$-competitive algorithm; more recently,~\cite{gan2026size} improved this to a constant factor competitive ratio for polynomial convex delays in the same uniform metric special case. For the bipartite variant on the line,  \cite{kuo2025online} gave a deterministic $\widetilde{O}(\sqrt{n})$-competitive algorithm, where $n$ is the number of requests on each side. Under independent Poisson arrival processes,~\cite{mari2025online} obtain constant competitiveness in expectation on arbitrary finite metrics.

\subsection{Organization}
Section~\ref{sec:prelims} introduces the notation and LP formulation for min-cost matching with linear delays. We present our algorithm in Section~\ref{sec:algorithm} and analyze it in Section~\ref{sec:linear-analysis}. Appendices~\ref{sec:concave} and~\ref{app:kway} extend the algorithm and analysis to concave delay costs and $k$-way matchings, respectively. Table~\ref{tab:notations} in the appendix summarizes the notation used throughout the paper.

% \subsubsection*{AI disclosure} TBA

\section{Background}
\label{sec:prelims}

Let $M_0 = (X, d_0)$ be a metric space.
In the \emph{Minimum Cost Perfect Matching with (Linear) Delay} ($\mpmd$) problem, the input is a finite set
$V = [n] = \{1,\ldots,n\}$ of requests, where $n\ge2$ is even.  A request $u$ arrives
at a finite time $a_u\ge0$ and is located at $x_u\in X$.  The algorithm knows neither $X$ nor the total number of requests $n$ in advance.

The goal of the algorithm is to produce a perfect matching in an online fashion. For any two requests $u$ and $v$, an algorithm  may irrevocably \emph{match}
them at any time 
$t \ge\max\{a_u,a_v\}$. Given the set $\{(u_i, v_i, t_i)\}_{i \in [n/2]}$ of requests matched by algorithm $\cA$, where $t_i$ is the time when
requests $u_i$ and $v_i$ are matched, its cost is defined as:
\[
  \cost(\cA) = \sum_{i=1}^{n/2} d_0(x_{u_i},x_{v_i})+(t_i - a_{u_i})+(t_i-a_{v_i}).
\]
The competitive ratio of an algorithm is defined to be the worst case ratio (over all inputs) of the cost of the algorithm compared to the cost of an optimal offline algorithm that knows the entire input ahead of time.

\paragraph{Time-expanded pseudometric.}
We can augment the metric space $M_0 = (X, d_0)$ by also adding time.  For the set $V$ of requests, consider the space $M = (V, d)$ where
\begin{equation}\label{eq:time-metric}
  d(u,v):=d_0(x_u,x_v)+\abs{a_u-a_v}.
\end{equation}
It can be easily checked that $M$ is a pseudometric. We shall repeatedly use the properties
\begin{equation}\label{eq:arrival-gap}
  d(u,v)\ge\abs{a_u-a_v}
  \quad\text{and}\quad
  d(u,v) \ge d_0(x_u,x_v) .
\end{equation}

The time-expanded view is convenient since the optimal solution to the online metric matching with delays problem is exactly the minimum weight perfect matching with the distance $d$.
\begin{lemma}[Lemma 1 of \cite{dufay2026deterministic}, \cite{bienkowski2018primal}, and \cite{azar2020deterministic}]\label{lem:offline-equivalence}
The cost of the optimal solution of $\mpmd$ equals the minimum weight of a perfect matching in $M$. 
\end{lemma}

\paragraph{Linear programming formulation.}
The time-expanded view also motivates the following linear programming (LP)  relaxation.
For $S\subseteq V$, let $\bd(S)$ be the set of edges with exactly one
endpoint in $S$.  The odd-cut relaxation (P) and its dual (D) are
\begin{align*}
\text{(P)}\qquad
 \min\quad &\sum_{e\in\binom V2}d(e)x_e,
 &
 \sum_{e\in\bd(S)}x_e&\ge1
 &&S\subseteq V, |S| \text{ odd} 
 &x_e&\ge0,\\[1ex]
\text{(D)}\qquad
 \max\quad &\sum_{S:\,\abs S\text{ odd}}y_S,
 &
 \sum_{S:\,e\in\bd(S)}y_S&\le d(e)
 &&\text{for every }e\in\binom V2,
 &y_S&\ge0.
\end{align*}
Lemma \ref{lem:offline-equivalence} and weak duality then yields:
\begin{lemma}
\label{lem:weak-duality}
The optimal value of the dual (D) is a lower bound on $\OPT$, the minimum weight of an offline perfect matching.
\end{lemma}

\section{Algorithm for $\mpmd$}
\label{sec:algorithm}
The algorithm has two distinct parts. The first part is a classical ball-growing procedure in the time-expanded metric space that constructs a feasible dual solution online. This procedure naturally partitions the requests into connected components defined by tight edge constraints of the dual. The second part of the algorithm then utilizes this component structure to determine which pairs of requests to match at each time.  By Lemma \ref{lem:weak-duality}, the dual solution from the first part provides a lower bound on $\OPT$.  

\paragraph{Slow dual growth}
Let $\alpha = \frac{1}{32}$ be a fixed constant\footnote{For simplicity of exposition, we will not attempt to optimize the constants.}. At any time, the algorithm maintains a partition of all requests that have arrived until that time into \emph{active components}. Formally, it maintains a mapping $\cC$ that assigns each arrived request to its active component. An active component is called \emph{growing} when its cardinality is odd and \emph{frozen} when its cardinality is even. Each growing active component $S$ is associated with a corresponding dual variable $y_S$. Let $y_S(t)$ denote the value of the dual variable $y_S$ at time $t$. 

When a new request $u$ arrives at time $a_u$, the singleton $\{u\}$ becomes an active \emph{growing} component and the corresponding dual variable is initialized to zero, i.e., $y_{\{u\}}(a_u) = 0$. The dual variables associated with each growing active component increase at a constant rate $\alpha$.
%as time progresses at unit rate.
The dual variables increase until an
edge $e = (u, v)$, with $u$ and $v$
belonging to different active 
components, becomes tight, i.e., 
$\sum_{S:\,e\in\bd(S)} y_S(t) = d(e)$; at that time, the two 
active components are \emph{merged}.   Formally, let $\cC(u)$ and $\cC(v)$ denote the active components containing $u$ and $v$ just before time $t$. Then we add $\cC(u) \cup \cC(v)$ as a new active component and the old components  $\cC(u)$ and $\cC(v)$ now become inactive. If multiple edges become tight at the same time, they are processed in arbitrary order. If the two end points of a tight edge are in the same active component when it is being processed, the edge gets skipped. For convenience, we define $\cC_t(u)$ to be the active component containing the  request $u$ at time $t$. Note that $\cC_t(u)$ is unique except at finitely many merge times.

We note that the dual solution constructed via this process is feasible since once a dual constraint becomes tight for some edge $e$ and the two endpoints merge into a single active component, we do not increase the dual variable for any set $S$ such that $e \in \bd(S)$. We further emphasize that the above dual growth procedure is only used to construct a feasible dual solution online and no actual matching decisions have been taken yet.

We formally define the following quantities that are useful for analysis. For uniformity of notation, we define $y_S(t) = 0$ whenever $S$ is even or has not been a growing active component by time $t$. For any time $t \geq 0$, 
\begin{align}
  L_{e}(t)&:=\sum_{S:\,e\in\bd(S)}y_S(t),
  &
  \reach(u,t)&:=\sum_{S\ni u}y_S(t),
  &
  \Dual(t)&:=\sum_Sy_S(t).
  \label{eq:load-reach-dual}
\end{align}
Thus, $L_e(t)$ is the total load on edge $e$ at time $t$, $\reach(u,t)$ is the total dual value accumulated around $u$, and $\Dual(t)$ is the total dual accumulated up to time $t$.
Note that in all three definitions, the sums include dual variables of components that are no longer active; such variables keep their values  
after their components merge.  
For convenience, for $t \leq 0$, we also define $y_S(t) = 0, \forall S \subseteq V$ and $\Dual(t) = 0$.
We also use $\Dual := \Dual(\infty)$ to be the value of the dual at the end of the algorithm.

We note that the above dual growth process depends only on the request arrivals and the growth rate $\alpha$, and is completely independent of the matching decisions. In Section \ref{sec:linear-analysis}, we bound the matching algorithm's cost by $O(\log n) \cdot \Dual$ and use weak duality to establish competitiveness. For such a charging argument to succeed, the online dual objective must itself be comparable to $\OPT$. In Appendix \ref{sec:fast-dual-counterexample}, we show that slowing down the dual growth is both necessary and sufficient for this property. If odd components grow at rate $1$, then we can have $\Dual \leq O(1/\sqrt{n}) \OPT$, whereas for any $\alpha \in (0,1)$, the online dual satisfies $\Dual \geq \frac{1-\alpha}{2} \cdot \OPT$.

\paragraph{Delayed best-effort matching}
 For every pair $(u, v)$ of requests define the \emph{maturation time} $\tau(u,v)$ as follows:
\begin{equation}\label{eq:deadline}
  \tau(u,v):=\min\{a_u, a_v\} + 2 d(u,v).
\end{equation}
We say that the request pair $(u,v)$ \emph{matures} at time $\tau(u,v)$. We first observe that the maturation time is necessarily after both the requests have arrived.  Indeed, suppose without loss of generality that $a_u \leq a_v$, then 
$\tau(u,v) - a_v = (a_v - a_u) + 2 d_0(x_u, x_v) \geq 0$. 

The algorithm arbitrarily matches pairs of requests $u, v$ that have reached their maturation time and are in the same dual component. Formally, at any time $t$ during the course of the algorithm, the algorithm matches requests $u$ and $v$ whenever $t \geq \tau(u,v)$, and $\cC_t(u) = \cC_t(v)$, and $u$ and $v$ are both unmatched. Ties are broken arbitrarily in case multiple request pairs become eligible at the same time.

We formalize the slow dual growth process in Algorithm \ref{alg:dual-growth} and the delayed best-effort matching construction in Algorithm~\ref{alg:delayed-matching}.

\paragraph{Time and event convention} We assume throughout the algorithm that time is continuous. The algorithm experiences finitely many \emph{events} where it potentially needs to take action. There are three kinds of events: (i) the arrival times of the requests, (ii) times when a dual constraint becomes tight for some edge, and (iii) the maturation times of pairs of unmatched requests. During any time between events, the algorithm simply increases the dual of each growing active component at rate $\alpha$. At event times, in case of ties we adopt the convention that newly arriving requests are processed first, followed by processing any tight edges (breaking ties among edges arbitrarily), and finally by matching matured request pairs (again breaking ties among pairs arbitrarily).

\begin{algorithm}[t]
\caption{Slow Dual Growth Process}
\label{alg:dual-growth}
\begin{algorithmic}
\State Fix a growth rate $\alpha = \frac{1}{32}$. Initially, there are no active components.
\While{time advances continuously}
    \State Every growing (i.e., odd) active component $S$ increases its dual variable $y_S$
    at rate $\alpha$. 
    \State Frozen (i.e., even) active components do not grow.
    % \State If a request $u$ arrives, create the singleton active component $\{u\}$ and set $y_{\{u\}}$ to zero.
    \If{request $u$ arrives}
    \State Create the singleton active component $\{u\}$ and set $y_{\{u\}}$ to zero.
    \EndIf
    \If{requests $u,v$
    in distinct active components become tight: $\displaystyle \sum_{S:\, |\{u,v\}\cap S|=1} y_S = d(u,v)$}
    \State Merge the respective active components into a single new active component.
    \State The two old components become
    inactive and their dual variables remain fixed thereafter.
    \EndIf
\EndWhile
\end{algorithmic}
\end{algorithm}

\begin{algorithm}[t]
\caption{Delayed Best-Effort Matching}
\label{alg:delayed-matching}
\begin{algorithmic}
\LongState{For every pair $(u, v)$ of requests, define its \emph{maturation time} by
\[
    \tau(u,v) := \min\{a_u,a_v\} + 2d(u,v).
\]}

\While{time $t$ advances continuously}
    \LongState{If $u$ and $v$ are unmatched at $t$, $\cC_t(u)=\cC_t(v)$, and $t \geq \tau(u,v)$, match $u$ to $v$.}
    \LongState{If several pairs become eligible at the same time $t$, choose among
    them arbitrarily and continue matching eligible pairs until none remain.}
\EndWhile
\end{algorithmic}
\end{algorithm}

\section{Competitive ratio analysis}
\label{sec:linear-analysis}

We first establish some simple observations that follow directly from the definition of the dual growth process.

\subsection{Structural properties}

Let $\mathcal{H}(t) \subseteq 2^V$ be the family of sets that have occurred as
active components by time $t$. We call components in $\mathcal{H}(t)$
\emph{historical components}. Whenever a tight edge $e$ triggers a merge of two active components,
we call $e$ a \textit{merge edge}. Let $F(t)$ be the graph consisting of all
merge edges added by time $t$. For $S \subseteq V$, let $F_S(t)$ denote the
subgraph of $F(t)$ induced by $S$. Let $U(t)$ denote the set of requests that are unmatched at time $t$. 

The following proposition lists some observations that follow directly from the definition of the dual growth process and the matching algorithm. We defer the proof to Appendix \ref{sec:omitted-proofs} for clarity.

\begin{restatable}{proposition}{basicproperties}
\label{prop:observations}
Let $t$ be an arbitrary time during the execution of the algorithm and $u$ be an arbitrary request that has arrived before time $t$. Then we have the following:
\begin{enumerate}
    \item \emph{(Laminarity)} The family $\cH(t)$ is \emph{laminar}, i.e., for any two distinct sets $S, T \in \cH(t)$, we have $S \cap T \in \{\emptyset, S, T\}$. The members of $\cH(t)$ that contain $u$ form a chain.
    \item  \emph{(Merge Trees)} For every historical component $S \in \cH(t)$, $F_S(t)$ is a tree.
    \item \emph{(Dual Feasibility)} The set $\{y_S(t)\}_{S \subseteq V : |S| \text{ odd}}$ of dual variables forms a feasible solution to the dual $(D)$.
    \item \emph{(Parity)} For every active component $A$ at any time $t$, we have  \[\abs{A\cap U(t)}\equiv\abs A\pmod 2.\]
    \item \emph{(Termination)} The algorithm terminates and always returns a perfect matching.
\end{enumerate}
\end{restatable}

For any request $u$, let $t_u$ be the time at which $u$ gets matched by the algorithm. Let $\Delay := \int_{0}^{\infty} \abs{U(t)} dt = \sum_{u \in V} (t_u - a_u)$ be the total delay cost incurred by the algorithm.

\begin{lemma}
\label{lem:redn-delay}
The total cost of the algorithm satisfies $\ALG \leq \frac{5}{4} \Delay$.    
\end{lemma}
\begin{proof}
    Let $\{(u_1, v_1, t_1), (u_2, v_2, t_2), \dots\}$ denote the matches made by the algorithm along with the time.   Let $\delta(u_i,v_i) = t_i - a_{u_i} + t_i - a_{v_i}$ be the contribution of $(u_i, v_i)$  to $\Delay$. Thus we have $\ALG = \sum_{i} d_0(x_{u_i}, x_{v_i}) + \delta(u_i, v_i)$ and $\Delay = \sum_i \delta(u_i, v_i)$. From \eqref{eq:deadline}, we have $t_i \geq \tau(u_i,v_i) = \min\{a_{u_i}, a_{v_i}\} + 2 d(u_i,v_i)$. So,
    \begin{align*}
        \delta(u_i, v_i) \geq 2 \min\{a_{u_i}, a_{v_i}\} + 4 d(u_i,v_i) - a_{u_i} - a_{v_i} = 4 d(u_i,v_i) - \abs{a_{u_i} - a_{v_i}} \geq 4 d_0(x_{u_i},x_{v_i}).
    \end{align*}
    Therefore, the total cost incurred by the algorithm for this pair is 
    \[d_0(x_{u_i}, x_{v_i}) + \delta(u_i, v_i) %\leq d(u_i,v_i) + \delta(u_i, v_i) 
    \leq \left(1 + \frac{1}{4}\right) \delta(u_i, v_i) = \frac{5}{4} \delta(u_i, v_i). \]
    The lemma follows by summing over all pairs.
\end{proof}

\subsection{From local dual charging to a global competitive ratio}
We will now identify a local charging property of the algorithm's delay cost against the accumulating dual that enables a bound on the final competitive ratio. 

Recall that $\Dual(t) := \sum_{S} y_S(t)$ is the total value of the dual accumulated by time $t$. For any request $u$ and time $t \geq a_u$, let $\age(u, t) := t - a_u$ be the amount of time $u$ has spent (either matched or unmatched) in the system by time $t$. The local charging theorem claims that at any time $t$ and age $r$, the total number of {\em unmatched} nodes of age $\Theta(r)$ is bounded by the growth in the dual objective in the previous $O(r)$ time interval. This is formalized as follows.

\begin{theorem}
\label{thm:recent-window}
For every $t\ge0$ and $r>0$, let $U(r,t):=\{u\in U(t):\age(u, t) \in[r, 2r)\}$ be the set of nodes that are unmatched at time $t$ and have age in $[r, 2r)$. Then
  \[r \cdot \abs{U(r,t)}
  \le32\bigl(\Dual(t)-\Dual(t-3r)\bigr).
\]
\end{theorem}

We prove Theorem~\ref{thm:recent-window} in Section~\ref{sec:local-argument}. Here we will show that it leads to a competitive ratio logarithmic in the number of requests.

Let $\wait(u) = t_u - a_u$ be the total delay incurred by request $u$. The following simple corollary of Theorem~\ref{thm:recent-window} shows that each request incurs only bounded delay.

\begin{corollary}
\label{corr:max-wait}
For any request $u$, $\wait(u) \leq 32 \cdot \Dual$. Consequently, for any time $t$, and for any $r > 32 \cdot \Dual$, we have $|U(r, t)| = 0$.
\end{corollary}
\begin{proof}
    Consider any time $t$ when the request $u$ is unmatched by the algorithm.  Then, applying Theorem~\ref{thm:recent-window} with $r = \age(u, t)$ yields the following:
    \begin{equation*}
        \age(u,t) \leq \age(u,t) \cdot |U(\age(u,t), t)| \leq 32(\Dual(t) - \Dual(t-3 \age(u,t))) \leq 32 \cdot \Dual.
    \end{equation*}
    The proof follows by considering $t$ that approaches $t_u$.
\end{proof}

\begin{lemma}
\label{lemma:window-integral}
    For every $L > 0$, 
    \[\int_{0}^\infty \Dual(t) - \Dual(t-L) dt = L \cdot \Dual. \]
\end{lemma}
\begin{proof}
    Let $t_{\mathrm{end}}$ denote the time when the dual growth procedure terminates and thus after which no dual variable changes. Fix $T := t_{\mathrm{end}} + L$, so that we have $\Dual(t) = \Dual, \forall t \geq T-L$.  Since $\Dual(t) = 0$
    for $t < 0$, applying a change of variables and telescoping yields
    \begin{equation*}
        \int_{0}^\infty \Dual(t) - \Dual(t-L) dt = \int_{0}^T \left(\Dual(t) - \Dual(t-L)\right) dt
        = \int_{T-L}^T \Dual(t) dt
        = L \cdot \Dual.  
        \qedhere
    \end{equation*}
\end{proof}

We now prove our main result: a logarithmic bound on the competitive ratio of Algorithm~\ref{alg:delayed-matching}.
% \manish{We could switch the definition of $U(r,s)$ to be the set of unmatched requests with age in $[r, er)$. That would get rid of all the $\frac{1}{\ln 2}$ terms that we need to carry around now. But it may not be worth it at the moment.}
\begin{theorem}
\label{thm:local-to-global}
    The Delayed Best-Effort Matching Algorithm (Algorithm~\ref{alg:delayed-matching}) achieves a competitive ratio of $\frac{40}{\ln 2}(1 + 3 \ln n)$.
\end{theorem}
\begin{proof}
    Consider an arbitrary request $u$ and any time $t$ such that $u$ is unmatched at time $t$.  We observe that this request belongs to the set $U(r, t)$ exactly for $r \in (\age(u, t)/2, \age(u, t)]$, and hence
    \begin{equation}
    \label{eq:lnint}
        \int_0^\infty \frac{1}{r} \cdot \mathbbm{1}[u\in U(r,t)] dr = \int_{\age(u, t)/2}^{\age(u, t)} \frac{1}{r} dr = \ln 2.
    \end{equation}
    Integrating \eqref{eq:lnint} over all time $t$,
    \[
        \int_0^\infty \int_0^\infty \frac{1}{r} \cdot \mathbbm{1}[u\in U(r,t)] dr\;dt  = \ln 2 \int_0^\infty \mathbbm{1}[u \in U(t)] dt = \ln2 \cdot \wait(u).   
    \]
    Summing over all requests $u$, then gives us an expression for the total delay.
    \begin{align}
      \Delay &= \sum_u \wait(u) = \frac{1}{\ln 2} \sum_{u} \int_0^\infty \int_0^\infty \frac{1}{r} \mathbbm{1}[u\in U(r,t)] dr\;dt \nonumber \\
      &= \frac{1}{\ln 2} \int_0^\infty \int_0^\infty \frac{1}{r} |U(r, t)| dr\;dt  = \frac{1}{\ln 2} \int_0^\infty \frac{1}{r} \int_0^\infty |U(r, t)| dt\;dr,
      \nonumber
    \intertext{
    where the change of the order of
    integration holds since the integrand is always non-negative. We split the outer integral to deal with very small values of $r$ separately; let $\rho$ be a fixed
    age threshold.}
    &= \frac{1}{\ln 2} \int_0^{\rho} \frac{1}{r} \int_0^\infty |U(r, t)| dt\; dr + \frac{1}{\ln 2} \int_{\rho}^{\infty} \frac{1}{r} \int_0^\infty |U(r, t)| dt\;dr.
    \label{eq:small-large}
\end{align}
We now bound these two terms in
\eqref{eq:small-large} separately.  Focusing on the first term, note that $\int_0^\infty |U(r, t)| dt = \sum_u \int_0^\infty \mathbbm{1}[u \in U(r, t)] dt$. For a fixed vertex $u$ and fixed age scale $r$, $u \in U(r, t)$ for at most $r$ time, and hence $\int_0^\infty \mathbbm{1}[u \in U(r, t)] dt \leq r$.  The first term hence
becomes
\begin{align}
\frac{1}{\ln 2} \int_0^{\rho} \frac{1}{r} \int_0^\infty |U(r, t)| dt\; dr 
& \leq \frac{1}{\ln 2} \int_0^{\rho} n dr = \frac{n\rho}{\ln 2}
\label{eq:small-large-small}
\end{align}
For the second term in \eqref{eq:small-large}, Corollary \ref{corr:max-wait} implies  $|U(r,t)| = 0$ for all $r > 32 \Dual$, and hence
\begin{align}
\frac{1}{\ln 2} \int_{ \rho}^{\infty} \frac{1}{r} \int_0^\infty |U(r, t)| dt\; dr 
& = \frac{1}{\ln 2} \int_{ \rho}^{32 \Dual} \frac{1}{r} \int_0^\infty |U(r, t)| dt\; dr \nonumber \\
    (\mbox{by Theorem~\ref{thm:recent-window}}) &\leq \frac{1}{\ln 2} \int_{\rho}^{32 \Dual}  \frac{1}{r} \int_0^\infty  \frac{1}{r} \cdot 32 \left(\Dual(t) - \Dual(t-3r)\right)  dt\; dr \nonumber \\
    (\mbox{by Lemma~\ref{lemma:window-integral}}) &= \frac{1}{\ln 2} \int_{\rho}^{32 \Dual} 96 \cdot \Dual \cdot \frac{1}{r} dr \nonumber \\
    & =  \frac{96 \cdot \Dual}{\ln 2} \int_{\rho}^{32 \Dual} \frac{1}{r} dr
    = \frac{96 \cdot \Dual}{\ln 2}\cdot \ln \frac{32 \cdot \Dual}{\rho}.
    \label{eq:small-large-large}
\end{align}
Setting $\rho = \frac{32 \cdot \Dual}{n}$ in \eqref{eq:small-large}, \eqref{eq:small-large-small}, 
and \eqref{eq:small-large-large} yields $\Delay \leq \frac{32 \cdot \Dual}{\ln 2}(1 + 3 \ln n)$. 
Applying Lemma \ref{lem:redn-delay}
and weak duality, we obtain 
$\ALG \leq \frac{5}{4} \Delay
\leq \frac{40}{\ln 2} (1 + 3 \ln n) \OPT$, completing the proof.
\end{proof}

\subsection{Local dual charging}\label{sec:local-dual-charging}
\label{sec:local-argument}

We will now prove the local charging argument, Theorem~\ref{thm:recent-window}. This charging forms the technical crux of the competitive analysis. At a high level, we decompose each active component into old versus recent (a.k.a. fresh) arrivals and show that each active component contains enough recent dual to pay for its unmatched requests. 

\subsubsection{Properties of the slow dual growth process}

Fix a time $t \ge 0$ and an age scale $r>0$; recall that $U(r, t)$ is the set of unmatched requests at time $t$ that have age in $[r, 2r)$.
We call a vertex $u$ \textit{old} if $a_u<t-3r$, and \textit{fresh} otherwise. Note that every vertex in $U(r,t)$ is fresh. Our goal is to prove the following per-component charging statement. Theorem~\ref{thm:recent-window} then follows by summing over all the active components.

\begin{lemma}
\label{lem:per-component-charging}
For every active component $C$ at time $t$, there is a family $\mathcal D
\subseteq
\{S\in\mathcal H(t): S\subseteq C\}$ of historical components such that 
\[
\sum_{S \in \mathcal D} y_S(t) 
\ge
\frac{r}{32}|C\cap U(r,t)|.
\]
Moreover, $y_S(t-3r)=0$ 
for every $S\in\mathcal D$.
\end{lemma}

The following three simple lemmas are a direct consequence of the slow dual growth process and form the foundations of our proof. Lemma \ref{lem:reach-bound} upper bounds the total dual growth that sets containing a single request can accumulate. Lemma \ref{lem:late-merger} formalizes the intuition that an edge becomes tight only after significant time has passed after the arrival of both its endpoints. Finally, Lemma \ref{lem:separation-of-unmatched-requests} states that if a dual component contains two unmatched requests, then they must be pairwise separated by a factor proportional to their age.

\begin{lemma}
\label{lem:reach-bound}
    For any request $u$ and time $t \geq a_u$, 
    \[\reach(u,t) :=\sum_{S\ni u}y_S(t) \leq \alpha(t - a_u).\]
\end{lemma}
\begin{proof}
    Request $u$ belongs to a single active component $S$ at any time in the interval $(a_u, t)$. 
    During this time interval, the dual variable grows at rate at most $\alpha$.  The proof follows. 
\end{proof}

\begin{lemma}
\label{lem:late-merger} Let $e=(u,v)$ be a merge edge with $a_u \leq a_v$, and let $\merge{e}$ be the time at which $e$ is added to the merge forest. Then we have \[\merge{e} \geq a_v + \left(\frac{1-\alpha}{2\alpha}\right) \cdot d(u, v). \]
\end{lemma}
\begin{proof}
    Before $v$ arrives at time $a_v$, no dual components containing $v$ have been created. Therefore, the load on $e$ at time $a_v$ is $L_e(a_v) = \reach(u, a_v) \leq \alpha(a_v - a_u) \leq \alpha \cdot d(u, v)$, by Lemma~\ref{lem:reach-bound}.  After time $a_v$, there is a remaining load of at least $(1-\alpha) d(u,v)$ on the edge to be covered. At most one dual component containing $u$ and at most one containing $v$ grow and load $e$, contributing a total rate of $2\alpha$; hence,
    $2 \alpha (\mu_e - a_v) \geq
    (1 - \alpha) d(u, v)$. 
    The proof follows.
\end{proof}

\begin{lemma}\label{lem:separation-of-unmatched-requests}
Let $u,v \in U(t)$ be distinct requests in the same active component at time
$t$. Then
\[
    d(u,v)
    >
    \frac{1}{2}
    \max\{t-a_u,\;t-a_v\}.
\]
\end{lemma}
\begin{proof} 
Since $u$ and $v$ are both unmatched, we must have $t<\tau(u,v)$. The proof then follows from the definition of $\tau(u,v)$.
%TBD. Suppose not true. Then the pair $(u,v)$ would already have passed its threshold for matching. Since the two requests are in the same active component, they would be eligible to match and could not both remain unmatched at time $s$.
\end{proof}

%We henceforth consider the remaining cases: either $C$ contains an old vertex, or $C$ contains no old vertices and $|C\cap U(r,s)|\ge 2.$

\subsubsection{Associating duals with fresh components}
\begin{figure}[t]
\centering
\resizebox{0.8\textwidth}{!}{
  %\documentclass[tikz,border=6pt]{standalone}
%\usepackage{amsmath}
%\usetikzlibrary{arrows.meta,calc}

%\begin{document}
\begin{tikzpicture}[
    x=0.9cm,y=0.9cm,
    line cap=round,line join=round,
    v/.style={circle,draw=black,fill=white,inner sep=0pt,minimum size=3.2pt,line width=.8pt},
    vr/.style={circle,draw=black,fill=white,inner sep=0pt,minimum size=4pt,line width=.9pt},
    hi/.style={circle,draw=violet,line width=2.5pt,inner sep=1.5pt},
    vm/.style={circle,draw=green!65!black,line width=2.5pt,inner sep=1.5pt},
    comp/.style={draw=red!65!black,line width=2.2pt},
    edge/.style={draw=black,line width=.8pt},
]

% Time levels
\def\xaxis{0}
\def\xright{12.1}
\def\ys{0}
\def\yone{1.6}
\def\ytwo{3.05}
\def\ythree{4.60}

% old region and highlighted age band
\fill[gray!18,rounded corners=10pt] (0.35,4.72) rectangle (12.0,8.0);
\fill[yellow!45,rounded corners=8pt] (0.35,1.68) rectangle (12.0,2.98);

% vertical time axis
\draw[-{Stealth[length=4mm]},line width=1.0pt] (0,7.9) -- (0,-1.0);
\node[rotate=90,anchor=south,font=\large] at (-1.0,5.9) {Time};

% horizontal reference lines
\draw[line width=.9pt] (-0.25,\ys) -- (12.15,\ys);
\draw[line width=.9pt] (-0.25,\yone) -- (12.15,\yone);
\draw[line width=.9pt] (-0.25,\ytwo) -- (12.15,\ytwo);
\draw[line width=.9pt] (-0.25,\ythree) -- (12.15,\ythree);

% labels on left
\node[anchor=e,font=\large] at (-0.35,\ys) {$t$};
\node[anchor=e,font=\large] at (-0.35,\yone) {$t-r$};
\node[anchor=e,font=\large] at (-0.35,\ytwo) {$t-2r$};
\node[anchor=e,font=\large] at (-0.35,\ythree) {$t-3r$};

% labels on right
\node[anchor=west,font=\large,align=left] at (12.15,-0.15) {current\\[-1mm] time};
\node[anchor=west,text=orange!85!black,font=\large] at (12.15,2.10) {$U(r,t)$};
\draw[-{Stealth[length=2.6mm]},orange!85!black,line width=1.0pt]
    (12.15,2.20) -- (11.1,2.6);

% old / fresh marker
\draw[-{Stealth[length=3mm]},line width=.9pt] (12.25,4.7) -- (12.25,5.7);
\draw[-{Stealth[length=3mm]},line width=.9pt] (12.25,4.45) -- (12.25,3.45);
\node[anchor=south,font=\large] at (12.75,5.1) {old};
\node[anchor=north,font=\large] at (12.95,4.15) {fresh};

% ---------------- black forest on the left ----------------
\coordinate (a1) at (0.9,3.72);
\coordinate (a2) at (1.28,5.25);
\coordinate (a3) at (2.45,5.95);
\coordinate (a4) at (4.35,7.05);
\coordinate (a5) at (2.95,5.35);
\coordinate (a6) at (2.28,3.78);
\coordinate (a7) at (3.42,3.55);
\coordinate (a8) at (4.62,3.75);
\coordinate (a9) at (6.24,4.98);
\coordinate (a10) at (7.22,3.55);
\coordinate (a11) at (6.75,2.15);

\draw[edge] (a1)--(a2)--(a3)--(a4);
\draw[edge] (a3)--(a5)--(a6)--(a7);
\draw[edge] (a9)--(a10)--(a11);

% ---------------- central/right forest ----------------
\coordinate (b1) at (7.55,7.65);
\coordinate (b2) at (9.65,7.35);
\coordinate (b3) at (8.08,5.52);
\coordinate (b4) at (10.78,5.62);
\coordinate (b5) at (6.86,3.62);
\coordinate (b6) at (7.78,2.18);
\coordinate (b7) at (9.38,3.78);
\coordinate (b8) at (10.72,3.48);
\coordinate (b9) at (11.00,2.68);
\coordinate (b10) at (9.02,2.05);
\coordinate (b11) at (8.42,0.95);
\coordinate (b12) at (9.22,0.85);
\coordinate (b15) at (10.30,2.78);
\coordinate (b16) at (10.05,2.08);
\coordinate (b17) at (10.43,1.95);

\draw[edge] (b1)--(b2);
\draw[edge] (b1)--(b3)--(b4);
\draw[edge] (b3)--(a10)--(b6);
\draw[edge] (b3)--(b7)--(b8)--(b9);
\draw[edge] (b7)--(b10);
\draw[edge] (b10)--(b11);
\draw[edge] (b10)--(b12);
\draw[edge] (b15)--(b16);
\draw[edge] (b15)--(b17);
\draw[edge] (b8)--(b15);

% ---------------- fresh component A (red overlay) ----------------
\coordinate (c1) at (1.78,1.20);
\coordinate (c2) at (2.42,1.92);
\coordinate (c3) at (1.92,2.25);
\coordinate (c4) at (2.83,2.78);
\coordinate (c5) at (4.25,2.58);
\coordinate (c6) at (5.10,3.80);
\coordinate (c7) at (5.70,2.66);
\coordinate (c8) at (4.15,0.72);

\draw[comp] (c1)--(c2)--(c3)--(c4)--(c5)--(c6);
\draw[comp] (c5)--(c7);
\draw[comp] (c4)--(c8);
\draw[comp] (a6)--(c3);
\draw[comp] (a6)--(a7);
\draw[edge] (a9)--(c6);

% draw all the nodes
\foreach \p in {a1,a2,a3,a4,a5,a9,a10,a11}
  \node[v] at (\p) {};
\foreach \p in {b1,b2,b3,b4,b6,b7,b8,b9,b10,b11,b12,b15,b16,b17}
  \node[v] at (\p) {};
\foreach \p in {c1,c2,c3,c4,c6,c7,c8,a6,a7}
  \node[vr] at (\p) {};
\foreach \p in {c5,b6,b16,b17}
  \node[vm] at (\p) {};

% highlighted vertices of A
\foreach \p in {c2,c3,c4,c7}
  \node[hi] at (\p) {};

% annotation for fresh component A
\node[align=center,text=red!70!black,font=\small] (freshlabel) at (6.72,0.75)
    {a fresh\\component\\$A$};
\draw[-{Stealth[length=2.6mm]},red!70!black,line width=1.0pt]
    (freshlabel.west) to[bend right=12] (3.95,1.12);

% Q_A annotation
\node[text=violet,font=\small] (qalabel) at (2.82,0.80) {$Q_A$};
\draw[-{Stealth[length=2.6mm]},violet,line width=1.0pt]
    (qalabel.north) -- (2.5,1.8);

\end{tikzpicture}
%\end{document}
}
\caption{Old and fresh nodes, and the merge tree $F_C(t)$, of a component $C$ at time $t$. Vertices with arrival times in the window $(t-2r, t-r]$ that are matched by time $t$ are depicted in green; the remaining vertices belong to the set $U(r,t)$. The thick red edges depict a fresh component $A$ and the purple nodes depict the vertex set $Q_A=U(r,t)\cap A$. }
\label{fig:matching-diagram}
\end{figure}
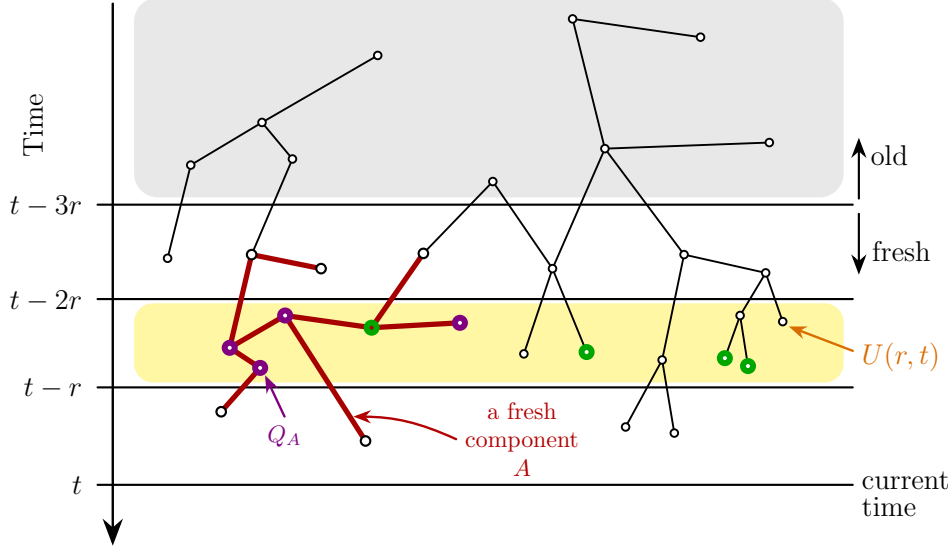

We first focus on the case when either the active component $C$ contains at least one old vertex, or  $|C \cap U(r,t)| \geq 2$; the remaining case of $C$ containing only fresh vertices and $|C \cap U(r,t)| = 1$ will be handled separately.
Consider an active component $C$ at time $t$. From its merge tree $F_C(t)$, remove every old vertex together with all incident edges. We call each connected component of the remaining forest a \textit{fresh component}; see Figure~\ref{fig:matching-diagram}. For any fresh component $A$, let $Q_A$ be the set of unmatched requests in $A$ with age in $[r, 2r)$ and let $m_A$ be its cardinality, i.e., 
\[
Q_A:=A\cap U(r,t),
\qquad
m_A:=|Q_A|.
\]
We only consider fresh components $A$ for which $m_A \geq 1$;
components with $m_A = 0$ require no charge. Finally, for any tree $T$ whose vertices are requests, define its length as
\[
\operatorname{len}(T):=\sum_{e\in E(T)}d(e).
\]

With a slight abuse of notation, let $F_A \subseteq F_C(t)$ be the subtree of the merge forest that is induced by component $A$. We first show that there exists a \emph{witness} tree $W_A \subseteq F_A$ such that total length of all edges in $W_A$ is at least $\Omega(m_A r)$. 

\begin{definition}[Witness Tree]
\label{def:witness-tree}
    Root the tree $F_C(t)$ at an arbitrary old vertex if one exists, otherwise root it arbitrarily. For any fresh component $A$, let $v_A \in A$ be the unique vertex in $A$ whose parent is old; if there are no such vertices, let $v_A$ be an arbitrary vertex in $A$. We define the \emph{witness tree} for fresh component $A$ to be the minimal subtree of $F_C(t)$ that spans all of $Q_A \cup \{v_A\}$.
\end{definition}

The key observation here is that Lemma \ref{lem:separation-of-unmatched-requests} enforces the constraint that every pair of requests in $Q_A$ is at distance greater than $\frac{r}{2}$ from each other. We prove the following simple lemma that shows that a set of vertices that are pairwise separated from each other forces a spanning tree covering those vertices to have a large total length.

\begin{lemma}
\label{lem:separated-terminals}
Let $k \geq 2$ and $\{u_1, \ldots, u_k\}$ be a set of vertices such that $d(u_i, u_j) > \rho,\ \forall i \ne j$. Let $T^*$ be any tree containing $u_1,\ldots,u_k$. Then
\[
\len(T^*)>\frac{k\rho}{2}.
\]
\end{lemma}

\begin{proof}
An Euler tour of the multigraph obtained by doubling each edge of $T^*$ traverses each edge exactly twice and so has length $2\len(T^*)$.
The tour visits the $k$ terminal vertices in some cyclic order and hence can be decomposed into $k$ consecutive walk segments. Without loss of generality, suppose the Euler tour visits $u_i$ before $u_j$ for $i < j$. Then, we have
\[2 \len(T^*) \ge \sum_{i = 1}^{k} d(u_i, u_{i+1}) > k \rho, \]
where the summation uses the convention that $u_{k+1} := u_1$.
\end{proof}

Lemma \ref{lem:separated-terminals} suffices to prove %Lemma \ref{lem:long-witness} 
that for any fresh component $A$ with $m_A \geq 2$, $W_A$ has size $\Omega(m_Ar)$, since we know by Lemma \ref{lem:separation-of-unmatched-requests} that each of the $m_A$ unmatched vertices are pairwise distance $> \frac{r}{2}$ away from each other. But the lemma cannot be applied directly for fresh components with $m_A = 1$. To deal effectively with such components, we show in Lemma \ref{lem:age-across-merge-edges} that the vertex $v_A$ chosen in Definition \ref{def:witness-tree} must be sufficiently old and hence serve as the second terminal so that Lemma \ref{lem:separated-terminals} may be applied.  

\begin{lemma}
\label{lem:age-across-merge-edges}
Let $e=(u,v)$ be a merge edge with $a_u\le a_v$, and let $\merge{e}$ be the time at which $e$ becomes tight. Then, for every $t\ge \merge{e}$,
\[
t-a_v
\ge
\frac{1-\alpha}{1+\alpha}(t-a_u).
\]
In particular, for $\alpha=1/32$, if $u$ is old and $v$ is fresh, then $t-a_v>\frac{31}{11}r>\frac{5}{2}r$, and $v\notin U(r,t)$.
\end{lemma}

\begin{proof}
Since \(t \ge \merge{e}\), Lemma \ref{lem:late-merger} gives
\[
t-a_v \ge \merge{e}-a_v \ge \frac{1-\alpha}{2\alpha} d(u,v) \implies d(u,v) \le \frac{2\alpha}{1-\alpha}(t-a_v).
\]
Hence,
$t-a_u
= (t-a_v) + (a_v-a_u) 
\le (t-a_v) + d(u,v) 
\le \left(1+\frac{2\alpha}{1-\alpha}\right)(t-a_v) 
= \frac{1+\alpha}{1-\alpha}(t-a_v).$
Rearranging gives the claim.
\end{proof}

We can then combine Lemmas \ref{lem:separated-terminals} and \ref{lem:age-across-merge-edges} to obtain the following lower bound on the length of the witness tree.

\begin{restatable}{lemma}{longwitnesslemma}
\label{lem:long-witness}
    Assume that either $C$ contains at least one old vertex, or $|C \cap U(r,t)| \geq 2$. For any fresh component $A \subseteq C$ with $m_A > 0$, the \emph{witness tree} $W_A \subseteq F_A$ satisfies $\len(W_A) > \frac{m_A r}{4}$.
\end{restatable}
\begin{proof}
    Suppose first that $C$ contains at least one old vertex. Then for each fresh component $A$, there exists a request $v_A \in A$ that is adjacent to an \emph{old} request. By Lemma \ref{lem:age-across-merge-edges}, we have $t - a_{v_A} > \frac{5}{2}r$ and hence $d(u, v_A) > \frac{r}{2}$ for each $u \in Q_A$. Thus we have a set of $m_A + 1$ vertices that are pairwise distance $>\frac{r}{2}$ away from each other (Lemma \ref{lem:separation-of-unmatched-requests}). Applying Lemma \ref{lem:separated-terminals} then shows that $\len(W_A) > \frac{(m_A + 1) r}{4}$, which is stronger than what we require.

    On the other hand, suppose $C$ has no old vertex. In this case, $C$ contains a single fresh component $A$ with $m_A \geq 2$. So we apply Lemma \ref{lem:separated-terminals} directly to obtain $\len(W_A) > \frac{m_A r}{4}$.
\end{proof}

We know that, by definition, any edge in the merge forest is tight, i.e., $\sum_{S \in \bd(e)} y_S = d(e)$. So one may hope that simply summing over all edges $e \in W_A$ should yield a total dual value of $\len(W_A) = \sum_{e \in W_A} d(e)$. However, this argument fails since the same set $S$ can cross many different edges of the tree $W_A$ and thus we may end up overcharging the same dual variable. 

To salvage this approach, we observe
the following. First, if we are merely trying to recover the cost of a path, we can recover half this cost from the duals crossing its edges, because any historical component $S$ can cross the path at most twice. We formalize this in Lemma \ref{lem:path-dual-bound} below.

\begin{figure}[t]
\centering
\resizebox{0.6\textwidth}{!}{
  %\documentclass[tikz,border=6pt]{standalone}
%\usepackage{amsmath}
%\usetikzlibrary{arrows.meta,calc}

%\begin{document}
\begin{tikzpicture}[
    x=0.9cm,y=0.9cm,
    line cap=round,line join=round,
    v/.style={circle,draw=black,fill=white,inner sep=0pt,minimum size=3.2pt,line width=.8pt},
    vr/.style={circle,draw=black,fill=white,inner sep=0pt,minimum size=4pt,line width=.9pt},
    hi/.style={circle,draw=violet,line width=2.5pt,inner sep=1.5pt},
    comp/.style={draw=red!65!black,line width=2.2pt},
    edge/.style={draw=black,line width=.8pt},
    compnew/.style={draw=green!65!black,line width=2.2pt},
]

% Time levels
\def\xaxis{0}
\def\xright{12.1}
\def\ys{0}
\def\yone{1.6}
\def\ytwo{3.05}
\def\ythree{4.60}

% old region and highlighted age band
\fill[gray!18,rounded corners=10pt] (0.35,4.72) rectangle (7.0,6.0);
\fill[yellow!45,rounded corners=8pt] (0.35,1.68) rectangle (7.0,2.98);

% vertical time axis
\draw[-{Stealth[length=4mm]},line width=1.0pt] (0,5.9) -- (0,-0.0);
\node[rotate=90,anchor=south,font=\small] at (-0.5,4.9) {Time};

% horizontal reference lines
\draw[line width=.9pt] (-0.25,\yone) -- (7.0,\yone);
\draw[line width=.9pt] (-0.25,\ytwo) -- (7.0,\ytwo);

% labels on left
\node[anchor=e,font=\small] at (-0.35,\yone) {$t-r$};
\node[anchor=e,font=\small] at (-0.35,\ytwo) {$t-2r$};

% labels on right
\node[anchor=west,text=orange!85!black,font=\small] at (5.8,2.10) {$U(r,t)$};
\node[anchor=west,text=green!65!black,font=\small] at (1.30,2.60) {$P_2$};
\node[anchor=west,text=red!65!black,font=\small] at (4.4,2.25) {$P_1$};
\node[anchor=west,text=black!65!black,font=\small] at (1.7,4.0) {$v_A$};

% old / fresh marker
\node[anchor=south,font=\small] at (5.7,5.5) {old vertices};

% ---------------- black forest on the left ----------------
\coordinate (a5) at (2.95,5.35);
\coordinate (a6) at (2.28,3.78);
\coordinate (a7) at (3.42,3.55);
\coordinate (a8) at (4.62,3.75);
\coordinate (a9) at (6.24,4.98);

% ---------------- fresh component A (red overlay) ----------------
\coordinate (c1) at (1.78,1.20);
\coordinate (c2) at (2.42,1.92);
\coordinate (c3) at (1.92,2.25);
\coordinate (c4) at (2.83,2.78);
\coordinate (c5) at (4.25,2.58);
\coordinate (c6) at (5.10,3.80);
\coordinate (c7) at (5.70,2.66);
\coordinate (c8) at (4.15,0.72);

\draw[edge] (c1)--(c2);
\draw[compnew] (c2)--(c3)--(a6);
\draw[comp] (c3)--(c4)--(c5)--(c7);
\draw[edge] (c5)--(c6);
\draw[edge] (c4)--(c8);
\draw[edge] (a6)--(a7);
\draw[edge,dashed] (a9)--(c6);
\draw[edge,dashed] (a5)--(a6);

% draw all the nodes
\foreach \p in {a5,a9}
  \node[v] at (\p) {};
\foreach \p in {c1,c2,c3,c4,c5,c6,c7,c8,a6,a7}
  \node[vr] at (\p) {};

% highlighted vertices of A
\foreach \p in {c2,c3,c4,c7}
  \node[hi] at (\p) {};

% Q_A annotation
\node[text=violet,font=\small] (qalabel) at (2.85,1.80) {$Q_A$};

\end{tikzpicture}
%\end{document}
}
\caption{Rooted decomposition of the fresh component $A$ from Figure~\ref{fig:matching-diagram}. The solid edges (red, green, and black) form the fresh component, and the dashed edges show its connections to old nodes. The witness tree $W_A$ spanning $Q_A\cup\{v_A\}$ is depicted with bold edges. This is decomposed into two paths---the red path $P_1$ and the green path $P_2$.}
\label{fig:path-decomp}
\end{figure}
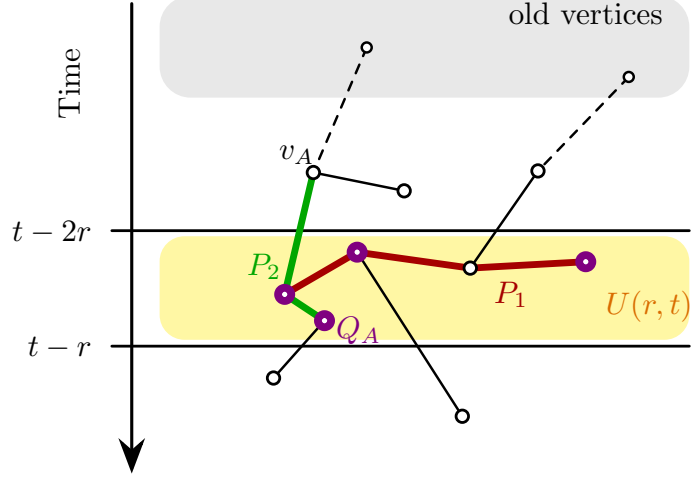

Second, if two paths share a vertex $b$, any historical component crossing both must also contain $b$. Excluding components containing $b$ from the charge for one of the paths prevents double counting across the two paths. We therefore decompose the tree $W_A$ into at most $m_A$ edge-disjoint paths; see Figure~\ref{fig:path-decomp}. We formalize this decomposition in Lemma \ref{lem:rooted-decomposition}. We then assign each path a family of historical components so that the families are pairwise disjoint and their total dual value is $\Omega(m_A r)$. Every historical component contributing to this bound contains a fresh vertex and hence satisfies $y_S(t-3r)=0$.

\begin{lemma}
\label{lem:path-dual-bound}
    Let $t \geq 0$ be an arbitrary time step. Let $P$ be a path of the merge forest $F(t)$ and $\len(P) = \sum_{e \in P} d(e)$. We say a set $S$ \emph{crosses}  path $P$ if $\exists e \in P, e \in \bd(S)$. Then
    \[\sum_{S : S \text{ crosses } P} y_S(t) \geq \frac{1}{2} \cdot \len(P)\]
\end{lemma}
\begin{proof}
    An edge $e$ is added to the merge forest exactly when its associated dual constraint becomes tight and the dual variables of sets that cross the edge do not change thereafter, i.e., $\sum_{S : e \in \bd(S)} y_S(t) = d(e),\ \forall e \in F(t)$. Summing over all edges in the path $P$ then gives
    \begin{equation*}
     \len(P) = \sum_{e \in P} d(e) = \sum_{e \in P} \sum_{S : e \in \bd(S)} y_S(t) = \sum_{S} y_S(t) |\bd_P(S)|,
    \end{equation*}
    where $\bd_P(S)$ denotes the set of edges of $P$ that have exactly one endpoint in $S$. Since any historical component $S$ with $y_S(t) > 0$ forms a connected subset of $F(t)$, we have $|\bd_P(S)| \leq 2$ and the lemma follows by rearranging.
\end{proof}

\begin{restatable}{lemma}{rooteddecompositionlemma}
\label{lem:rooted-decomposition}
There exist edge-disjoint paths $P_1,\ldots,P_q$ that cover $W_A$, with $q\le m_A$. Further we can associate a set family $\mathcal{D}_j$ with each path $P_j$ such that 
\begin{enumerate}
    \item $\mathcal{D}_1, \ldots, \mathcal{D}_q$ are pairwise disjoint,
    \item $\forall S \in \mathcal{D}_A$, we have $v_A \notin S$,
    \item $\displaystyle \sum_{S\in \mathcal{D}_A}y_S(t)
\ge
\left(\frac12-\alpha\right)\operatorname{len}(W_A)
-2\alpha q r > \frac{m_A r}{32}$,
\end{enumerate}
where $\mathcal{D}_A := \bigcup_{j=1}^q \mathcal{D}_j$.
\end{restatable}
\begin{proof}
    Consider any fixed fresh component $A$. Since $W_A$ is chosen to be the minimal subtree of $F_C(t)$ that spans all of $Q_A \cup \{v_A\}$, all leaves of $W_A$ (except possibly $v_A$) are nodes in $Q_A$. We root $W_A$ at $v_A$ and thus the number of non-root leaves is at most $m_A = |Q_A|$. 

    Let $\ell_1, \ldots \ell_q$ be the (non-root) leaf nodes of $W_A$. We associate a path with each leaf via the following iterative procedure: Initialize $T = W_A$. For each leaf $\ell_i$ in order, let $P_i$ be the maximal path starting from $\ell_i$ such that all its interior vertices are of degree two and are not the root vertex. Let $b_i$ be the other endpoint of $P_i$; we call $b_i$ as the \emph{anchor} of path $P_i$. Note that either $b_i$ has degree at least three, or $b_i$ is the root. The tree $T$ is then modified by deleting all edges from $P_i$. Note that by construction, we have that the paths $P_1, \ldots, P_q$ are edge-disjoint and together cover the entire tree $W_A$. Further $q \leq m_A$ since each path begins at a unique non-root leaf node. See Figure~\ref{fig:path-decomp} for an example.    

    For each $1 \leq j \leq q$, we define a family of sets as follows: 
    \[
\mathcal{D}_j
:=
\{S\in\mathcal{H}(t):
y_S(t)>0,\;
S\text{ crosses }P_j,\;
b_j\notin S\}.
\]

By construction, for any $i < j$, the path $P_i$ can share at most the single anchor node $b_i$ with $P_j$. Since each $S \in \cH(t)$ is connected in $F_C(t)$ and none of the sets in $\mathcal{D}_i$ contains $b_i$ by definition, we must have that the families $\mathcal{D}_i$ and $\mathcal{D}_j$ are disjoint. Also, by construction, the node $v_A$ is chosen to be the root and hence cannot be in the interior of any path $P_j$. Thus for any $j$, either $v_A \notin P_j$ or $v_A = b_j$, and in either case cannot be included in any set in $\cD_j$. 

Consider any family $\cD_j$. We now show that $\sum_{S \in \cD_j} y_S(t) \geq (\frac{1}{2} - \alpha) \len(P_j) - 2\alpha r$. The lemma then follows by summing over all the $q \leq m_A$ families. Since $\cD_j$ contains all sets that cross path $P_j$ except those that contain $b_j$, we have:
\begin{equation}
\label{eq:peeling-reach-bound}
    \sum_{S \in \cD_j} y_S(t) \ge \sum_{S : S \text{ crosses } P_j} y_S(t) - \reach(b_j, t) \geq \frac{1}{2} \len(P_j) - \reach(b_j, t),
\end{equation}
where the last inequality uses Lemma \ref{lem:path-dual-bound}. Suppose $u_j \in U(r, t)$ is the leaf node forming the other endpoint of $P_j$. To bound the last term, note that Lemma \ref{lem:reach-bound} gives 
$\reach(b_j, t) \leq \alpha(t - a_{b_j})$. If $a_{u_j} \leq a_{b_j}$, then $t-a_{b_j} \leq t - a_{u_j} \leq 2r$ and substituting above yields the claim. Suppose otherwise, then 
\[\reach(b_j, t) \leq \alpha(t - a_{b_j}) = \alpha(t - a_{u_j} + a_{u_j} - a_{b_j}) \leq \alpha (2r + d(u_j, b_j)) \leq \alpha(2r + \len(P_j))\]
where the last step used the triangle inequality. Substituting into \eqref{eq:peeling-reach-bound} yields the claim.
\end{proof}

Combining Lemmas \ref{lem:long-witness} and \ref{lem:rooted-decomposition} suffices to show that the total dual is at least $\Omega(m_A r)$ for any fresh component $A$. To complete the proof, we still need to argue that two different fresh components (within the active component $C$) do not charge to the same dual variable. Once again we use the fact that historical components correspond to connected subtrees of $F_C(t)$. Our choice of $v_A$ in Definition \ref{def:witness-tree} implies that any historical dual component that intersects two different fresh components $A$ and $A'$ must contain either $v_A$ or $v_{A'}$. We therefore have the following. 

\begin{restatable}{lemma}{disjointfreshcomponentslemma}
\label{lem:disjoint-fresh-component-charges}
For distinct fresh components $A$ and $A'$,
$\cD_A \cap \cD_{A'} = \emptyset$.
\end{restatable}
\begin{proof}
    Suppose for contradiction that there exists a set $S \in \cD_A \cap \cD_{A'}$. Since $S$ crosses a path in the witness tree $W_A$, $S$ contains at least one vertex $u_A \in A$. Similarly, $S$ contains at least one vertex $u_{A'} \in A'$. By Proposition~\ref{prop:observations}, $F_S(t)$ is connected and hence $S$ contains all vertices on the path joining $u_A$ and $u_{A'}$. However, by construction, such a path must contain at least one of $v_A$ or $v_{A'}$. In either case, that contradicts Lemma \ref{lem:rooted-decomposition}, part(ii).
\end{proof}

We now handle the remaining corner case. Suppose $C$ contains no old vertices and $|C \cap U(r,t)|=1$. The following baseline bound then immediately gives the desired result. Since $C$ contains no old vertices, all sets $S \subseteq C$ satisfy $y_S(t - 3r) = 0$ since none of these sets could have existed as active components before $t-3r$.

\begin{restatable}{lemma}{baselinepaymentlemma}
\label{lem:baseline-payment}
Let $u\in U(r,t)$ and let $C$ be its active component at time $t$. Then
\[
\sum_{S\subseteq C} y_S(t)\ge \frac{r}{32}.
\]
\end{restatable}

\begin{proof}
Let $\{u\} = S_1 \subset S_2 \subset \dots \subset C$ denote the chain of active components containing request $u$ since its arrival until time $t$. If all of these components are odd (except possibly at event times that are of zero measure), then since odd active components grow at rate $\alpha$, we have $\sum_{S \subseteq C; u \in S} y_S(t) = \alpha(t - a_u) \geq \alpha r = \frac{r}{32}$ as desired.

Suppose $u$ belongs to an even component for a non-trivial interval of time and let $\sigma \in (a_u, t]$ be the last time when $u$ was in an even component, say $S_j$. Then by Proposition \ref{prop:observations}, $S_j$ must contain another unmatched request $v$. Applying Lemma \ref{lem:separation-of-unmatched-requests}, we have $d(u, v) \geq \frac{\sigma - a_u}{2}$. Consider the path $P_{uv}$ joining requests $u$ and $v$ in the merge tree $F_{S_j}(\sigma)$. Applying Lemma \ref{lem:path-dual-bound} and the  triangle inequality, we get
\begin{equation*}
    \sum_{S \subseteq C} y_S(\sigma) \geq \frac{1}{2}\sum_{e \in P_{uv}} d(e) \geq \frac{d(u,v)}{2} \geq \frac{\sigma-a_u}{4}.
\end{equation*}
After time $\sigma$, the component containing $u$ is always odd and so the total dual value of components containing $u$ increases by $\alpha(t - \sigma)$. Adding these two contributions together, 
\begin{equation*}
    \sum_{S \subseteq C} y_S(t) \geq \frac{\sigma-a_u}{4} + \alpha(t - \sigma) > \alpha(t - a_u) \geq \frac{r}{32}, 
\end{equation*}
where the second inequality uses $\alpha = \frac{1}{32} < \frac{1}{4}$.
\end{proof}

We now use these to complete the 
proof of the per-component charging lemma.
\begin{proof}[Proof of Lemma \ref{lem:per-component-charging}]
    First, let us suppose that $C$ contains no old vertices and thus we have a single fresh component $A = C$. If $m_A = 1$, then the lemma follows directly from Lemma \ref{lem:baseline-payment}. If $m_A \geq 2$, then the lemma follows from Lemma \ref{lem:rooted-decomposition}.

    Now, suppose $C$ contains an old vertex. Since the sets $Q_A$ partition $C \cap U(r, t)$, we have $\sum_{\text{fresh components } A \subset C} m_A = |C \cap U(r, t)|$. By Lemma \ref{lem:disjoint-fresh-component-charges}, the families $\mathcal{D}_A$ are disjoint, so defining $\mathcal{D} = \bigcup_A \mathcal{D}_A$ and summing over Lemma \ref{lem:rooted-decomposition} then gives
    \[\sum_{S \in \mathcal{D}} y_S(t) > \frac{r}{32} \sum_{A} m_A = \frac{r}{32} |C \cap U(r, t)|\]
    Moreover since each $S \in \mathcal{D}$ contains a fresh vertex, it satisfies $y_S(t-3r) = 0$.
\end{proof}

\paragraph{AI methodology.}
All core algorithmic and technical ideas in this paper --- including the slowed dual growth process, the delayed matching rule, and the local dual charging analysis --- were developed entirely by the human authors. We used GPT 5.6 Sol, Claude Opus 5, and Gemini 3.1 Pro to extend the framework to concave delay functions and $k$-way matchings, as well as to assist with proofreading, exposition and presentation improvements, and independent verification of the technical proofs across the manuscript. The authors have thoroughly verified the correctness and originality of all content, including all mathematical claims and bibliographic references, and take full responsibility for the submission.

\bibliographystyle{myplainnat}
\bibliography{references}

\appendix
\begin{table}[htbp]
\centering
\small
\caption{Table of Notations.}
\label{tab:notations}
\begin{tabular}{@{}ll@{}}
\toprule
\textbf{Notation} & \textbf{Description} \\
\midrule
\multicolumn{2}{@{}l}{\textit{Problem \& Metric Space}} \\
$M_0 = (X, d_0)$ & Underlying metric space with point set $X$ and metric $d_0$ \\
$V = \{1, \ldots, n\}$ & Set of $n$ requests ($n \ge 2$, $n$ is even) \\
$a_u$ & Arrival time of request $u \in V$ ($a_u \ge 0$) \\
$x_u$ & Spatial location of request $u$ in $(X, d_0)$ \\
$M = (V, d)$ & Time-expanded pseudometric space over requests $V$ \\
$d(u, v)$ & Time-expanded distance: $d_0(x_u, x_v) + |a_u - a_v|$ \\
$\OPT$ &  Optimal offline MPMD cost \\
$\operatorname{cost}(\mathcal{A})$ & Total cost of algorithm $\mathcal{A}$ \\
\midrule
\multicolumn{2}{@{}l}{\textit{Dual Growth \& Linear Program}} \\
$\partial(S)$ & Cut of $S \subseteq V$; edges with exactly one endpoint in $S$ \\
$x_e$ & Primal LP variable for edge $e \in \binom{V}{2}$ \\
$y_S, y_S(t)$ & Dual variable for odd set $S \subseteq V$ (at time $t$) \\
$\alpha$ & Dual growth rate constant ($\alpha = 1/32$) \\
$L_e(t)$ & Load on edge $e$ at time $t$: $\sum_{S: e \in \partial(S)} y_S(t)$ \\
$\operatorname{reach}(u, t)$ & Dual accumulated around request $u$ by time $t$: $\sum_{S \ni u} y_S(t)$ \\
$\Dual(t), \Dual$ & Dual objective at time $t$, and final dual value $\Dual(\infty)$ \\
\midrule
\multicolumn{2}{@{}l}{\textit{Algorithm State \& Matching}} \\
$\mathcal{C}_t(u), \mathcal{C}(u)$ & Active component containing request $u$ at time $t$ \\
$\tau(u, v)$ & Maturation time of pair $(u, v)$: $\min\{a_u, a_v\} + 2d(u, v)$ \\
$\mathcal{H}(t)$ & Laminar family of historical active components up to time $t$ \\
$F(t), F_S(t)$ & Forest of merge edges added by time $t$, and subgraph induced by $S$ \\
$\merge{e}$ & Event time when merge edge $e$ becomes tight \\
$U(t)$ & Set of requests that remain unmatched at time $t$ \\
$t_u$ & Time at which request $u$ is matched \\
$\operatorname{wait}(u)$ & Delay incurred by request $u$: $t_u - a_u$ \\
$\Delay$ & Total delay cost: $\int_0^\infty |U(t)|\,dt = \sum_u \operatorname{wait}(u)$ \\
\midrule
\multicolumn{2}{@{}l}{\textit{Local Charging Analysis}} \\
$\operatorname{age}(u, t)$ & Age of request $u$ at time $t$: $t - a_u$ \\
$r$ & Age scale parameter ($r > 0$) \\
$U(r, t)$ & Unmatched requests at time $t$ with age in $[r, 2r)$ \\
$A$ & Fresh component of active component $C$ \\
$Q_A, m_A$ & Set $A \cap U(r, t)$ and its cardinality $|Q_A|$ \\
$\operatorname{len}(T)$ & Sum of edge lengths in subtree $T$: $\sum_{e \in E(T)} d(e)$ \\
$v_A$ & Fresh boundary vertex of $A$ adjacent to an old component \\
$W_A$ & Minimal subtree of $F_C(t)$ spanning $Q_A \cup \{v_A\}$ \\
$P_j, b_j, u_j$ & Edge-disjoint path in $W_A$ with anchor $b_j$ and terminal $u_j \in Q_A$ \\
$\mathcal{D}_j, \mathcal{D}_A$ & Family of sets crossing $P_j$ avoiding $b_j$, and $\mathcal{D}_A = \bigcup_j \mathcal{D}_j$ \\
\bottomrule
\end{tabular}
\end{table}

\newpage

\section{Proofs from Section \ref{sec:linear-analysis}}
\label{sec:omitted-proofs}

\basicproperties*
\begin{proof}
    We establish the five properties in order.
    \begin{enumerate}
        \item \emph{(Laminarity)} At any time, the family of currently active components is pairwise disjoint. A new active component is formed either as a singleton $\{v\}$ when a request $v$ arrives, or as the union $C_1 \cup C_2$ of two disjoint active components $C_1$ and $C_2$ that immediately become inactive. By induction on the sequence of events up to time $t$, any two sets $S, T \in \mathcal{H}(t)$ are either disjoint or one is a subset of the other. Consequently, the members of $\mathcal{H}(t)$ that contain $u$ are pairwise comparable under set inclusion and therefore form a chain.

        \item \emph{(Merge Trees)} We proceed by induction on the creation of historical components. When a request $v$ arrives, $F_{\{v\}}$ consists of a single vertex and no edges, which is a tree. When two distinct active components $C_1$ and $C_2$ merge at time $s \le t$ via a tight edge $e = (v, w)$ with $v \in C_1$ and $w \in C_2$, the edge $e$ joins the two vertex-disjoint trees $F_{C_1}(s)$ and $F_{C_2}(s)$, so $F_{C_1 \cup C_2}(s)$ is a tree. Moreover, because merge edges are only added between vertices in distinct active components, no edge with both endpoints in $S \in \mathcal{H}(t)$ is ever added after $S$ is formed. Thus $F_S(t)$ remains a tree.

        \item \emph{(Dual Feasibility)} By definition, $y_S(t) \ge 0$ for all $S \subseteq V$ and $y_S(t) = 0$ whenever $|S|$ is even. Consider any edge $e = (v, w) \in \binom{V}{2}$ with $a_v \le a_w$. For $t < a_v$, we have $L_e(t) = 0$. For $t \in [a_v, a_w]$, since $w$ has not yet arrived, only active components containing $v$ can cross $e$. Because $v$ belongs to a single active component at any time and its dual variable grows at rate at most $\alpha$, we have
    \[
        L_e(t) = \mathrm{reach}(v, t) \le \alpha(t - a_v) \le \alpha(a_w - a_v) \le \alpha d(e) \le d(e).
    \]
    After time $a_w$, the load $L_e(t)$ increases continuously only while $v$ and $w$ lie in distinct active components. Once $L_e(t) = d(e)$ (or earlier if $v$ and $w$ merge via other tight edges), $v$ and $w$ belong to the same active component, after which no active component $S$ satisfies $e \in \partial(S)$ and $L_e$ remains fixed thereafter. Hence $L_e(t) \le d(e)$ for all $t \ge 0$.

    \item \emph{(Parity)} The congruence $|A \cap U(t)| \equiv |A| \pmod 2$ holds whenever a singleton component $A = \{v\}$ is created upon the arrival of $v$. Merging two active components $A_1$ and $A_2$ preserves the congruence since
    \[
        |(A_1 \cup A_2) \cap U(t)| = |A_1 \cap U(t)| + |A_2 \cap U(t)| \equiv |A_1| + |A_2| = |A_1 \cup A_2| \pmod 2,
    \]
    and matching a pair of requests in $A$ removes exactly two requests from $A \cap U(t)$, which also preserves parity.
    \item \emph{(Termination)} After all $n$ requests have arrived, any odd active component continues growing its dual variable at rate $\alpha > 0$. Since $n$ is even, the number of odd active components is always even, so an odd active component cannot occur alone. Thus, as long as an odd active component exists, a dual constraint between distinct active components eventually becomes tight and triggers a merge. Since at most $n - 1$ merges can occur, eventually every active component has even size and dual growth terminates. The parity invariant then implies that any active component containing unmatched requests must contain at least two of them. Since all maturation times $\tau(v, w)$ are finite, the algorithm continues matching eligible pairs until no unmatched requests remain. \qedhere
    \end{enumerate}
\end{proof}

\section{Metric matching with concave delays}
\label{sec:concave}

In this section we show that our algorithm and analysis naturally extend to the more general setting where the delay experienced by any request is specified by an arbitrary concave function.

\subsection{Background}

Let $\phi: \bbR_{\geq 0} \rightarrow \bbR_{\geq 0}$ be a continuous, increasing, unbounded, and concave function with $\phi(0) = 0$ that represents the delay incurred. \footnote{Both the concavity and unboundedness assumptions are necessary to achieve a sublinear competitive ratio. \cite{liu2018impatient} show a linear lower bound on the competitive ratio of every deterministic algorithm for convex delay functions, even when the underlying metric is uniform. For unboundedness, note that if the delay function is bounded by $B$ and the algorithm receives two vertices at distance $kB$, then the algorithm must eventually match these vertices at some finite point of time, incurring a cost of at least $kB$. But the adversary can then generate a new request at each location, achieving an optimal cost of at most $2B$. Therefore, the competitive ratio of any algorithm is unbounded.}
Let $\phi^{-1}$ be its inverse (which is guaranteed to exist). We also assume for simplicity that the derivative $\phi'$ always exists, although this assumption is not strictly necessary.
We use the notation and problem set up as in Section~\ref{sec:prelims}, except that if requests $u$ and $v$ are matched at time $t$, the algorithm incurs a cost of $d_0(x_u, x_v) + \phi(t - a_u) + \phi(t - a_v)$.

The time-expanded metric with concave delays is defined as follows.  Let $M_\phi = (V, d_{\phi})$ where
\[d_\phi(u, v) = d_0(x_u, x_v) + \phi(|a_u - a_v|).\]
$M_\phi$ is a valid (pseudo)metric. Indeed, consider any three requests $u, v, w \in V$. Since $\phi$ is concave, we have $\phi(|a_u - a_w|) \leq \phi(|a_u - a_v|) + \phi(|a_v - a_w|)$. Hence it follows that $d_\phi(u, w) \leq d_\phi(u,v) + d_\phi(v,w)$.

Since the optimal offline solution for metric matching with concave delays continues to be the optimal minimum weight perfect matching in the metric $M_\phi$, the LP formulation in Section \ref{sec:prelims} holds (by replacing $d$ with $d_\phi$).

\subsection{Algorithm}
\label{sec:concave-algorithm}

The algorithm is almost identical to that described in Section \ref{sec:algorithm} for linear delays. We only describe the differences here. The algorithm design continues to have the two distinct parts of slow dual growth and a delayed matching.

\paragraph{Slow dual growth}
Unlike in the linear delays case, each request in the growing, active component incurs delay at a \emph{different} rate in the concave delays case. In particular, since the delay function is concave, old requests accrue delay at a slower rate than newer requests.

We maintain the same slowed dual growing procedure as in the linear case and increase the dual variables of active components with an odd number of requests. For any odd component $S$, the dual variable $y_S$ increases at rate $\alpha$ times the rate of delay incurred by the oldest request in $S$. Formally, suppose an odd component $S$ remains active over the time interval $[\sigma_1, \sigma_2)$, then its dual increases by $\alpha\cdot(\phi(\sigma_2 - a_u) - \phi(\sigma_1 - a_u))$ where $a_u$ is the arrival time of the oldest request in set $S$. 

\paragraph{Delayed best-effort matching}
We modify the maturation time of a request pair to \[\tau_{\phi}(u,v) = \min\{a_u, a_v\} + \phi^{-1}(2 d_\phi(u,v)).\] The matching rule is otherwise unchanged, so the algorithm matches any pair of requests $u, v$ that both belong to the same dual component and have passed their maturation time.

\subsection{Analysis}
\label{sec:concave-analysis}

The analysis for general concave delays closely mirrors that for the linear case. We show that the local dual charging theorem continues to hold with $\phi(r)$ replacing $r$ in the bound on $\abs {U(r,t)}$. The modified local dual charging then lends itself to a global competitive ratio bound in a similar fashion by scaling the integral with the rate of dual growth. Since a large part of the analysis carries over from the linear case, we primarily highlight just the differences below.

The terms $L_{e}(t) := \sum_{S : e \in \bd(S)} y_S(t)$, $\reach(u, t) := \sum_{S \ni u} y_S(t)$, $\Dual(t) := \sum_{S} y_S(t)$ are defined exactly as in the linear case. Since the structure of both the dual growth and matching remain similar, Proposition \ref{prop:observations} continues to hold in the concave case.  The following are the analogues of Lemmas \ref{lem:reach-bound} and \ref{lem:separation-of-unmatched-requests} and are useful to bound the dual contribution of any single request and the distance between any pair of unmatched requests respectively.
\begin{lemma}
\label{lem:concave-reach-bound}
    For any request $u$ and time $t \geq a_u$, 
    \[\reach(u,t) :=\sum_{S\ni u}y_S(t) \leq \alpha \cdot \phi(t - a_u).\]
\end{lemma}
\begin{proof}
    At any time $\theta \in (a_u, t)$, the request $u$ belongs to a single active component $S$. If the active component $S$ is even, then the $y_S(\theta)$ does not increase at all. On the other hand, if $S$ is odd, $y_S(\theta)$ grows at rate $\alpha \phi'(\theta - a_f) \leq \alpha \phi'(\theta-a_u)$ where $f$ is the earliest request in $S$. Integrating, $\reach(u,t) \leq \int_{a_u}^t \alpha \phi'(\theta-a_u) d\theta = \alpha \phi(t-a_u)$.
\end{proof}

\begin{lemma}
\label{lem:concave-separation-of-unmatched-requests}
Let $u,v \in U(t)$ be distinct requests in the same active component at time
$t$. Then
\[
    d_\phi(u,v)
    >
    \frac{1}{2}
    \max\{\phi(t-a_u),\; \phi(t-a_v)\}.
\]
\end{lemma}

\begin{proof}
Suppose for contradiction that $d_{\phi}(u, v) \leq \frac{1}{2}
    \max\{\phi(t-a_u),\; \phi(t-a_v)\}$. Rearranging gives $t \geq \min\{a_u, a_v\} + \phi^{-1}(2d_{\phi}(u,v)) = \tau_\phi(u,v)$. Thus at time $t \geq \tau_\phi(u,v)$, the pair $u, v$ belongs to the same active component and thus is eligible for matching. But this contradicts that $u, v$ are unmatched at time $t$.
\end{proof}

For any request $u$, let $t_u$ be the time at which $u$ gets matched by the algorithm. Let $\Delay_\phi :=  \sum_{u \in V} \phi(t_u - a_u)$ be the total delay cost incurred by the algorithm.
\begin{lemma}
\label{lem:concave-redn-delay}
The total cost of the algorithm satisfies $\ALG \leq \frac{5}{4} \Delay_\phi$.    
\end{lemma}
\begin{proof}
    Suppose requests $u_i, v_i$ are matched by the algorithm at time $t_i$ and without loss of generality suppose that $a_{u_i} \leq a_{v_i}$. Let $\delta(u_i,v_i) = \phi(t_i - a_{u_i}) + \phi(t_i - a_{v_i})$ be the contribution of this pair to $\Delay_{\phi}$. We have $t_i \geq \tau_{\phi}(u_i,v_i) = a_{u_i} + \phi^{-1}(2 d_\phi(u_i,v_i))$. So,
    \begin{align*}
        \delta(u_i, v_i) \geq 2 d_\phi(u_i, v_i) + \phi\left(\phi^{-1}(2d_\phi(u_i, v_i)) -(a_{v_i} - a_{u_i})\right) \geq 4 d_\phi(u_i,v_i) - \phi(a_{v_i} - a_{u_i}) \geq 4 d_0(x_{u_i},x_{v_i}),
    \end{align*}
    where the second inequality used the concavity of $\phi$. Therefore, the
    total cost incurred by the algorithm for this pair is 
    \[d_0(x_{u_i}, x_{v_i}) + \delta(u_i, v_i) %\leq d(u_i,v_i) + \delta(u_i, v_i) 
    \leq \left(1 + \frac{1}{4}\right) \delta(u_i, v_i) = \frac{5}{4}  \delta(u_i, v_i). \]
    The proof follows by summing over all pairs.
\end{proof}

\subsubsection{Local dual charging}

We now show that the following analogue of the local dual charging theorem holds even for concave delays. The rest of the section is dedicated to the proof of the following.

\begin{theorem}
\label{thm:concave-recent-window}
For every $t\ge0$ and $r>0$, let $U(r,t):=\{u\in U(t):\age(u, t) \in[r, 2r)\}$ be the set of nodes that are unmatched at time $t$ and have age in $[r, 2r)$. Then
  \[\phi(r) \abs{U(r,t)}
  \le34 \cdot \bigl(\Dual(t)-\Dual(t-3r)\bigr).
\]
\end{theorem}

Fix a time $t \geq 0$ and an age scale $r > 0$, and an active component $C$ at time $t$. Our goal is to show the following analogue of Lemma \ref{lem:per-component-charging}: There exists a set of historical components $\cD \subset\{S \in \cH(t) : S \subseteq C\}$ such that $\sum_{S \in \cD} y_S(t) \geq \frac{\phi(r)}{34} |C \cap U(r,t)|$. Further $y_S(t  -3r) = 0$ for all $S \in \cD$. Theorem \ref{thm:concave-recent-window} then follows by summing over all active components. Henceforth, we focus our attention on a single active component $C$.

The slight degradation in the constant factor (32 to 34) arises from a weaker baseline payment. The following  handles the corner case when $C$ has no request that is older than $3r$ and has a single vertex in $U(r, t)$.
\begin{lemma} 
\label{lem:concave-baseline}
Let $u \in U(r, t)$ and $C$ be its active component at time $t$. Then
    \[\sum_{S \subseteq C} y_S(t) \geq \frac{\phi(r)}{34}. \]
\end{lemma}

\begin{proof}
    Let $\{u\} = S_1 \subset S_2 \subset \cdots \subset C$ denote the chain of active components containing $u$ since its arrival time until time $t$. Let $\sigma \in [a_u, t]$ be the last time that $u$ was in an even component. Assume that $\sigma = a_u$ if there is no such time.

    After time $\sigma$, since $u$ is always in some odd component, the dual of the corresponding component grows. Let $o$ denote the oldest request in $C$. Now at any time $\theta \in (\sigma, t)$, the dual variable of the component containing $u$ at time $\theta$ grows at rate at least $\alpha \phi'(\theta - a_o)$. Thus, overall, the dual increases by at least  $\alpha(\phi(t - a_o) - \phi(\sigma - a_o)) \geq \alpha(\phi(t - a_o) - \phi(\sigma - a_u) - \phi(a_u - a_o)) \geq \alpha(\phi(t - a_u)- \phi(\sigma - a_u) - \phi(a_u - a_o))$.
    
    If $\sigma > a_u$, then at time $\sigma$, we have another unmatched vertex $v$, with $d_\phi(u, v) \geq \phi(\sigma - a_u)/2$ (Lemma \ref{lem:concave-separation-of-unmatched-requests}), so the path bound (Lemma \ref{lem:path-dual-bound}) gives a dual of at least $\phi(\sigma - a_u)/4 > \alpha \phi(\sigma-a_u)$.
    
    Combining, we get a total dual of $\alpha (\phi(t-a_u) - \phi(a_u - a_o)) \geq \alpha (\phi(r) - \phi(a_u - a_o))$. It remains to upper bound the last term, $\phi(a_u - a_o)$. We have $\phi(a_u - a_o) \leq d_\phi(u, o) \leq 2 \sum_{S \subseteq C} y_S(t)$ (Lemma \ref{lem:path-dual-bound}). Substituting and rearranging gives $\sum_{S \subseteq C} y_S(t) \geq \frac{1}{34} \phi(t - a_u) \geq \frac{\phi(r)}{34}$ as desired.
\end{proof}
The proof of the local charging theorem for the other cases (when $C$ contains at least one old request, or if $|C \cap U(r,t)| \geq 2$) is essentially identical to the linear case. We provide a proof of one lemma that needs a slight modification and a proof sketch for the rest of the argument. The following lemma shows that even with concave delays, for any merge edge $(u, v)$, the ages of the two endpoints remain close to each other.
\begin{lemma} 
\label{lem:concave-age-across-merge-edge}
If $e=(u,v)$ is a merge edge with $a_u \leq a_v$ that is merged at time $\merge{e}$, for any $t \geq \merge{e}$,
\[t - a_v \geq \frac{1-\alpha}{1+\alpha} (t - a_u)\]
\end{lemma}
\begin{proof}
    At time $a_v$, we have $L_e(a_v) = \reach(u, a_v) \leq \alpha d_\phi(e)$ (Lemma \ref{lem:concave-reach-bound}). At any time $\theta \geq a_v$, the load $L_e$ increases at rate at most $2\alpha \phi'(\theta-a_v)$. So by time $\merge{e}$, each endpoint contributes at most $\alpha \phi(\merge{e} - a_v)$ load. Since the edge becomes tight at time $\merge{e}$, we get
    \[\phi(t - a_v) \geq \phi(\merge{e} - a_v) \geq \frac{1-\alpha}{2\alpha} d_\phi(e) \geq \frac{1-\alpha}{2\alpha} \cdot \phi(a_v - a_u).\]
    But note that since $\phi$ is concave and non-decreasing, for any $\beta > 1$ and $x,y \geq 0$, we have $\phi(x) \geq \beta \phi(y) \implies x \geq \beta y$. 
    Substituting above, we get $t - a_v \geq \frac{1-\alpha}{2\alpha} \cdot (a_v - a_u) = \frac{1-\alpha}{2\alpha} ((t - a_u) - (t - a_v))$. Rearranging gives the claim.
\end{proof}
We now prove the per-component local charging bound for general concave delay functions.
\begin{lemma}
\label{lem:concave-per-component-charging}
For every active component $C$ at time $t$, there is a family $\mathcal D
\subseteq
\{S\in\mathcal H(t): S\subseteq C\}$ of historical components such that 
\[
\sum_{S \in \mathcal D} y_S(t) 
\ge
\frac{\phi(r)}{34}|C\cap U(r,t)|.
\]
Moreover, $y_S(t-3r)=0$ for every $S\in\mathcal D.$
\end{lemma}

\begin{proof}[Proof Sketch]
    First, let us suppose that $C$ contains no old vertices and thus we have a single fresh component $A = C$. If $m_A = 1$, then the proof follows directly from Lemma \ref{lem:concave-baseline}.

    In the other cases, Lemma \ref{lem:concave-age-across-merge-edge} and Lemma \ref{lem:concave-separation-of-unmatched-requests} imply that we can find a witness tree $W_A$ such that the pairwise distance between any pair of terminals is at least $\frac{\phi(r)}{2}$, yielding
    $\len(W_A)>m_A\phi(r)/4$.  We can then perform the rooted decomposition as in Lemma \ref{lem:rooted-decomposition}. For each path $P_j$ with endpoints $u_j\in Q_A$ and $b_j$,  Lemma~\ref{lem:concave-reach-bound} and concavity yield $\reach(b_j,t)\le\alpha(2\phi(r)+\len(P_j))$.  Hence,
\[
\sum_{S\in D_A}y_S(t)
\ge \left(\frac12-\alpha\right)\len(W_A)
    -2\alpha m_A\phi(r)
> \frac{7}{128}m_A\phi(r)
> \frac{m_A\phi(r)}{34}.
\]
As in Lemma \ref{lem:disjoint-fresh-component-charges}, the charge families are disjoint, and every
charged dual variable vanishes at time $t-3r$.  Summing over the fresh components then completes the proof.
\end{proof}

\subsubsection{Local to global}
Finally, we show that integrating the local dual charging theorem over both the age scales and time give the desired logarithmic bound on the competitive ratio.  We note that Lemma \ref{lemma:window-integral} holds in this setting with no change. 
Let $t_u$ be the time when request $u$ is matched by the algorithm and let $\wait(u) = \phi(t_u - a_u)$ be the delay incurred by request $u$.  The following gives an upper bound on the age of any request during the course of the algorithm. 

\begin{corollary}
\label{corr:concave-max-wait}
For any request $u$, $\wait(u) \leq 34 \cdot \Dual$. Consequently, for any time $t$, and for any $r > \phi^{-1}(34 \cdot \Dual)$, we have $|U(r, t)| = 0$.
\end{corollary}
\begin{proof}
    Consider any time $t$ when the request $u$ is unmatched by the algorithm. Then, applying Theorem~\ref{thm:concave-recent-window} with $r = \age(u, t)$ yields the following:
    \begin{equation*}
        \phi(\age(u,t)) \leq \phi(\age(u,t)) \cdot |U(\age(u,t), t)| \leq 34(\Dual(t) - \Dual(t-3 \age(u,t))) \leq 34 \cdot \Dual.
    \end{equation*}
    The proof follows by considering $t$ that approaches $t_u$.
\end{proof}

\begin{theorem}
    The Delayed Best-Effort Matching Algorithm (Section \ref{sec:concave-algorithm}) achieves a competitive ratio of $\frac{85}{2 \ln 2} \left(1 + 3 \ln n\right)$ for min-cost perfect matching with concave delays.
\end{theorem}
\begin{proof}
Consider an arbitrary request $u$ and any time $t$ such that $u$ is unmatched at time $t$ and $\age(u,t)>0$.
    We observe that this request belongs to the set $U(r,t)$ exactly for
    $r \in (\age(u,t)/2,\age(u,t)]$, and hence
    \begin{equation}
    \label{eq:lnint-concave}
        \int_0^\infty \frac{\phi'(r)}{r}
        \mathbbm{1}[u\in U(r,t)]\,dr
        = \int_{\age(u,t)/2}^{\age(u,t)}
        \frac{\phi'(r)}{r}\,dr
        \geq \phi'(\age(u,t)) \cdot \ln 2,
    \end{equation}
    where the inequality follows from concavity of $\phi$.
    Integrating \eqref{eq:lnint-concave} over all time $t$,
    \[
        \int_0^\infty \int_0^\infty
        \frac{\phi'(r)}{r}\mathbbm{1}[u\in U(r,t)]\,dr\,dt
        \geq \ln 2 \int_{a_u}^{t_u}\phi'(t-a_u)\,dt
        = \ln 2 \cdot \wait(u).
    \]
    Summing over all requests $u$ then gives us an upper bound
    on the total delay:
    \begin{align}
        \Delay_\phi
        &= \sum_u \wait(u)
        \leq \frac{1}{\ln 2}\sum_u
        \int_0^\infty\int_0^\infty
        \frac{\phi'(r)}{r}\mathbbm{1}[u\in U(r,t)]\,dr\,dt
        \nonumber\\
        &= \frac{1}{\ln 2}\int_0^\infty
        \frac{\phi'(r)}{r}\int_0^\infty |U(r,t)|\,dt\,dr,
        \nonumber
    \intertext{
    where the change of the order of integration holds since the
    integrand is non-negative. We split the outer integral to deal
    with very small values of $r$ separately; let $\rho>0$ be a
    fixed age threshold.}
        \Delay_\phi
        &\leq \frac{1}{\ln 2}\int_0^\rho
        \frac{\phi'(r)}{r}\int_0^\infty |U(r,t)|\,dt\,dr
        + \frac{1}{\ln 2}\int_\rho^\infty
        \frac{\phi'(r)}{r}\int_0^\infty |U(r,t)|\,dt\,dr.
        \label{eq:small-large-concave}
    \end{align}
    We now bound these two terms separately. For a fixed request
    $u$ and fixed age scale $r$, $u\in U(r,t)$ for at most $r$
    time, and hence
    $\int_0^\infty |U(r,t)|\,dt\leq nr$.
    The first term therefore satisfies
    \begin{equation}
    \label{eq:small-concave}
        \frac{1}{\ln 2}\int_0^\rho
        \frac{\phi'(r)}{r}\int_0^\infty |U(r,t)|\,dt\,dr
        \leq \frac{n}{\ln 2}\int_0^\rho \phi'(r)\,dr
        = \frac{n\phi(\rho)}{\ln 2}.
    \end{equation}
    For the second term, Corollary~\ref{corr:concave-max-wait} implies
    $|U(r,t)|=0$ for all $r>\phi^{-1}(34\Dual)$.
    Thus, applying Theorem~\ref{thm:concave-recent-window} and Lemma~\ref{lemma:window-integral},
    \begin{align}
        \frac{1}{\ln 2}\int_\rho^\infty
        \frac{\phi'(r)}{r}\int_0^\infty |U(r,t)|\,dt\,dr
        &= \frac{1}{\ln 2}\int_\rho^{\phi^{-1}(34\Dual)}
        \frac{\phi'(r)}{r}\int_0^\infty |U(r,t)|\,dt\,dr
        \nonumber\\
        &\leq \frac{34}{\ln 2}
        \int_\rho^{\phi^{-1}(34\Dual)}
        \frac{\phi'(r)}{r\phi(r)}
        \int_0^\infty
        \left(\Dual(t)-\Dual(t-3r)\right)\,dt\,dr
        \nonumber\\
        &= \frac{102\Dual}{\ln 2}
        \int_\rho^{\phi^{-1}(34\Dual)}
        \frac{\phi'(r)}{\phi(r)}\,dr
        = \frac{102\Dual}{\ln 2}
        \ln\frac{34\Dual}{\phi(\rho)}.
        \label{eq:large-concave}
    \end{align}
    Setting $\rho=\phi^{-1}(34\Dual/n)$ in
    \eqref{eq:small-large-concave}, \eqref{eq:small-concave},
    and \eqref{eq:large-concave} yields
    \[
        \Delay_\phi \leq
        \frac{34 \cdot \Dual}{\ln 2}(1+3\ln n).
    \]
    Applying Lemma~\ref{lem:concave-redn-delay} and weak duality, we obtain
    \[
        \ALG\leq \frac54 \Delay_\phi
        \leq \frac{85}{2\ln 2}(1+3\ln n)\OPT,
    \]
    completing the proof.
\end{proof}

\section[k-way matchings]{$k$-way matchings}
\label{app:kway}

In this section, we consider the $k$-way matching with delays problem and show that our algorithm and analysis naturally extends to this setting with only minor modifications. 

% We take the sum of pairwise distances as the group connection cost and
% obtain an $O(k\log n)$-competitive algorithm. We then obtain competitive
% ratios of $O(k^2\log n)$ for diameter costs and $O(k^3\log n)$ for
% $H$-metric costs by comparison with the sum objective. Finally, a direct
% dual-feasibility argument gives an $O(k\log n)$-competitive ratio for
% minimum spanning tree costs using the same algorithm.

\subsection{Background}
\label{app:kway:background}

Motivated by applications to online multiplayer gaming, the $k$-way
matching with delays problem seeks to partition requests into groups of
size $k$ while minimizing their total delay and the connection cost of
each group. 

Defining the cost of a group of $k$ requests involves extending the concept of a metric space to give distances to subsets of $k$ elements (rather than just pairwise distances). \citet{melnyk2021kway} introduce the notion of
$H$-metrics, formally defined in Section~\ref{app:kway:hmetrics}, and
give an $O(k^5\log N)$-competitive algorithm, where $N$ is the number of
points in the metric space. Informally, an $H$-metric is defined by a ground set $X$ of elements and a function $H : X^k \rightarrow \mathbb{R}_{\ge 0}$ that satisfies a generalization of the triangle inequality. \citet{kakimura2025kway} give a
primal-dual algorithm with competitive ratio $O(k^2 n)$, where $n$ is the
number of requests, without requiring knowledge of the metric space in
advance.

We take a slightly different approach and instead study $k$-way matching with delays when the underlying requests are embedded in a typical pairwise metric space $(X, d_0)$. 
%We first consider requests in an ordinary pairwise metric space $(X,d_0)$. 
Fix an integer $k\geq 2$, known to the algorithm, and assume
that the total number of requests $n\geq k$ is finite and divisible by
$k$. As before, neither $n$ nor the metric space needs to be known in
advance. A request $u$ arrives at location $x_u$ at time $a_u\geq 0$.
The algorithm can irrevocably group any $k$ unmatched requests at any
time after all requests in the group have arrived. For a group $G$, let
\[
  \operatorname{sum}(G)
    := \sum_{\{u,v\}\subseteq G} d_0(x_u,x_v),
\]
where the sum is over unordered pairs of distinct requests. Forming
$G$ at time $t$ incurs cost
\[
  \operatorname{sum}(G)+\sum_{u\in G}(t-a_u).
\]
For $k=2$, this is exactly the MPMD objective from Section~\ref{sec:prelims}.

Our main result is an $O(k\log n)$-competitive algorithm for this
sum objective. In Section~\ref{app:kway:other-costs}, we extend this result to the diameter, $H$-metric, and minimum spanning tree objectives. More generally, our algorithm and analysis immediately extend to any group connection cost $c(G)$ that is sandwiched by the sum of pairwise distances: if $\lambda \sumcost(G) \le c(G) \le \Lambda \sumcost(G)$ for all groups $G$ of size $k$, then our algorithm yields an $O(\frac{\Lambda}{\lambda} \cdot k \log n)$-competitive guarantee. 
All of these results use linear delay costs and require no advance
knowledge of the underlying metric space.

\subsection{Linear program}
\label{app:kway:lp}

For a group $G$ of $k$ requests, write
\[
  a_G^{\max}:=\max_{v\in G}a_v,
  \qquad
  W(G):=\sum_{u\in G}(a_G^{\max}-a_u).
\]
This is the minimum delay cost of forming $G$. Let $V$ be the set of
all requests. As in the pairwise relaxation, we impose cut constraints;
here, the relevant sets are those whose cardinalities are not divisible
by $k$. Every partition of $V$ into $k$-element groups contains a group
crossing each such set. Hence the following is a relaxation of the
offline problem:
\[
\begin{array}{lll}
(P^k) & \displaystyle\min
  \sum_{G\subseteq V:\,|G|=k}
  \bigl(\operatorname{sum}(G)+W(G)\bigr)x_G & \\[2mm]
& \displaystyle\text{s.t.}\quad
  \sum_{G:\,|G|=k,\,0<|S\cap G|<k} x_G\geq 1
  & \text{for every }S\subseteq V\text{ with }k\nmid |S|,\\[2mm]
& x_G\geq 0
  & \text{for every }G\subseteq V\text{ with }|G|=k.
\end{array}
\]
Its dual is
\[
\begin{array}{lll}
(D^k) & \displaystyle\max\sum_{S:\,k\nmid |S|}y_S & \\[2mm]
& \displaystyle\text{s.t.}\quad
  \sum_{\substack{S:\,k\nmid |S|\\0<|S\cap G|<k}} y_S
  \leq \operatorname{sum}(G)+W(G)
  & \text{for every }G\subseteq V\text{ with }|G|=k,\\[2mm]
& y_S\geq 0
  & \text{for every }S\subseteq V\text{ with }k\nmid |S|.
\end{array}
\]
By weak duality, every feasible solution to $(D^k)$ lower bounds the
offline optimum.

\subsection{Algorithm}
\label{app:kway:algorithm}

The algorithm follows the framework of Section~\ref{sec:algorithm}. An active component
$S$ is growing when $k\nmid |S|$ and frozen otherwise, and each growing
component increases its dual variable at rate $\alpha$. For technical reasons, we use $\alpha=1/64$ instead of $\alpha=1/32$ here.
Matching decisions involve $k$ requests rather than two. To account for
the sum objective, we use the auxiliary time-expanded distance
\[
  \widehat d(u,v):=(k-1)d_0(x_u,x_v)+|a_u-a_v|
\]
in both the dual process and the maturation rule. This is the usual
time-expanded distance for the spatial metric $(k-1)d_0$; arrival times
and delay costs are unchanged. Importantly, the dual growth continues to be restricted by the {\em pairwise} constraints. 
Specifically, for every $u,v\in V$, the dual growth process enforces
\begin{equation}
  \sum_{S:\,|S\cap\{u,v\}|=1}y_S\leq\widehat d(u,v).
  \label{app:kway:pair-constraints}
\end{equation}
As we show below, the factor $k-1$ in the auxiliary time-expanded metric ensures that satisfying the pairwise constraints implies the group constraints for sum costs. Whenever a pair in distinct active components becomes tight, we merge
those components and add the tight edge to the merge forest. Historical
dual variables retain their values after a merge. Matched requests
remain in their components; only their unmatched status changes.
We set $y_S=0$ for every set with $k\mid |S|$ and every set that has not
been a growing active component.

For each group $G$ of $k$ requests, define
\begin{equation}
  \Delta(G):=\max_{u,v\in G}\widehat d(u,v),
  \qquad
  \tau(G):=\min_{u\in G}a_u+2\Delta(G).
  \label{app:kway:maturation}
\end{equation}
The algorithm forms a group $G$ whenever all its requests are unmatched, lie in
the same active component, and $t\geq\tau(G)$. Notice that
$\Delta(G)\geq\max_{u\in G}a_u-\min_{u\in G}a_u$, so maturation
cannot precede the arrival of any request in $G$.

% We use the event convention of Section~3: process arrivals first, then
% tight edges until no tight edge joins distinct components, and finally
% eligible groups until none remain. Fix deterministic rules for all
% ties. The state at time $t$ means the state after processing these
% events. There are finitely many arrivals, merges, and group maturation
% times. A direct implementation can enumerate the $k$-element groups;
% its running time is polynomial in $n$ for each fixed $k$.

\subsection{Analysis}
\label{app:kway:analysis}

Once again, the analysis closely mirrors that in Section~\ref{sec:linear-analysis}. 
%The largest distinction is that since the algorithm still restricts dual growth using pairwise constraints, we need a way to relate pairwise constraints to the $k$-way dual variables. The following lemma shows that these pairwise constraints still provide a $k$-way lower bound after losing a factor of $k/2$. Let $\{y_S\}_{S : k \nmid |S|}$ denote the dual variables obtained by the above dual growth process and let $\Dual = \sum_S y_S$ denote the total dual accumulated.
Let $U(t)$ be the unmatched requests at time $t$, $\{y_S\}_{S : k \nmid |S|}$ denote the dual variables obtained by the above dual growth process, and write
\[
  \Dual(t):=\sum_S y_S(t),
  \qquad
  \Dual:=\Dual(\infty).
\]
As before, set $y_S(t)=\Dual(t)=0$ for $t\leq 0$. 
We first show that the dual $\{y_S\}_{S : k \nmid |S|}$ is feasible for $(D^k)$.

%The sum objective allows us to pass from pairwise constraints to group constraints without scaling the dual variables.

\begin{lemma}
\label{app:kway:sum-dual}
Every nonnegative vector $\{y_S\}_{S:\,k\nmid |S|}$ satisfying
\eqref{app:kway:pair-constraints} is feasible for $(D^k)$.
Consequently, $\Dual\leq\OPT$ for sum costs.
\end{lemma}
\begin{proof}
Fix a group $G$ of size $k$. A set $S$ with $0<|S\cap G|<k$
separates exactly $b(k-b)$ pairs in $G$, where $b=|S\cap G|$.
Since $b(k-b)\geq k-1$, summing the pairwise constraints gives
\[
  (k-1)\sum_{S:\,0<|S\cap G|<k}y_S \leq \sum_{\{u,v\}\subseteq G}\sum_{S:\,|S\cap \{u,v\}|=1}y_S
    \leq\sum_{\{u,v\}\subseteq G}\widehat d(u,v).
\]
For each pair, $|a_u-a_v|\leq(a_G^{\max}-a_u)+(a_G^{\max}-a_v)$.
Each request occurs in $k-1$ unordered pairs, and hence
\[
  \sum_{\{u,v\}\subseteq G}\widehat d(u,v)
    \leq(k-1)\operatorname{sum}(G)+(k-1)W(G).
\]
Dividing by $k-1$ proves the group constraint. Weak duality gives the
last assertion.
\end{proof}

All merge-edge and path lengths in this section are measured using $\widehat d$. The structural arguments from Section~\ref{sec:linear-analysis} apply to the
spatial metric $(k-1)d_0$. In particular, the laminarity and merge tree properties of Proposition~\ref{prop:observations},
as well as Lemmas~\ref{lem:reach-bound}, \ref{lem:late-merger},
\ref{lem:age-across-merge-edges}, and \ref{lem:path-dual-bound} continue to hold, with $\alpha=1/64$. In particular, an arrived request $u$ satisfies
\begin{equation}
  \reach(u,t):=\sum_{S\ni u}y_S(t)
    \leq\alpha(t-a_u),
  \label{app:kway:reach}
\end{equation}
exactly as in Lemma~\ref{lem:reach-bound}. Thus, when a new request $v$ arrives, the load on its constraint with any earlier request $u$ is at most $\alpha(a_v-a_u)\leq\widehat d(u,v)$. 

We now briefly describe how the parity invariant of Proposition~\ref{prop:observations} is replaced by a modulo-$k$ invariant, and justify why the algorithm terminates and satisfies $\ALG\le \frac32 \Delay$, where $\Delay$ is its total delay cost. 

%; in particular, the laminarity and merge-tree arguments of Proposition~\ref{prop:observations} are unchanged. The pairwise constraints also remain feasible. Indeed, an arrived request $u$ satisfies
%\begin{equation}
%  \reach(u,t):=\sum_{S\ni u}y_S(t)
%    \leq\alpha(t-a_u),
%  \label{app:kway:reach}
%\end{equation}
%exactly as in Lemma~\ref{lem:reach-bound}. Thus, when a new request $v$ arrives, the load on its constraint with any earlier request $u$ is at most $\alpha(a_v-a_u)\leq\widehat d(u,v)$. Constraints between arrived requests are enforced by merging, and constraints involving a future request are feasible by the same bound. None of these arguments uses parity.

\begin{lemma}
\label{app:kway:structure-cost}
For every active component $C$ at time $t$,
\[
  |C\cap U(t)|\equiv |C|\pmod{k}.
\]
The algorithm terminates and produces a partition into groups of size
$k$. For every group $G$ formed at time $t$,
\begin{equation}
  \sum_{u\in G}(t-a_u)\geq(k+1)\Delta(G).
  \label{app:kway:group-delay}
\end{equation}
In particular, for sum costs,
$\ALG \leq\tfrac32\Delay$, where
$\Delay:=\int_0^\infty |U(t)|\,dt$.
\end{lemma}
\begin{proof}
The congruence holds initially and is preserved by arrivals, merges,
and the removal of exactly $k$ unmatched requests from a component.
After the last arrival, a growing component cannot be the only active
component, since $k\mid n$. If a growing component exists, its dual
eventually makes a constraint to another component tight, causing a
merge. There are at most $n-1$ merges, so after a finite time all active
components are frozen. Each then contains a multiple of $k$ unmatched
requests. All group maturation times are finite, so all remaining
requests are eventually grouped.

Now let $u^\star$ be an oldest request in a group $G$ formed at time
$t$, and set $\Delta=\Delta(G)$. By the maturation rule,
$t-a_{u^\star}\geq 2\Delta$. Also,
$a_u-a_{u^\star}\leq\widehat d(u,u^\star)\leq\Delta$
for each $u\in G$.
Therefore
\[
  \sum_{u\in G}(t-a_u)
    =k(t-a_{u^\star})
       -\sum_{u\in G\setminus\{u^\star\}}(a_u-a_{u^\star})
    \geq 2k\Delta-(k-1)\Delta=(k+1)\Delta.
\]
Since $d_0(x_u,x_v)\leq\widehat d(u,v)/(k-1)$, the connection
cost satisfies
\[
  \operatorname{sum}(G)
    \leq\frac{1}{k-1}\binom{k}{2}\Delta
    =\frac{k}{2}\Delta
    \leq\frac{k}{2(k+1)}\sum_{u\in G}(t-a_u).
\]
Summing over the groups proves
$\ALG\leq(1+k/(2(k+1)))\Delay
\leq\tfrac32\Delay$.
\end{proof}

\subsubsection{Local dual charging}
\label{app:kway:local}

For the local dual charging argument, we define fresh components as in Section \ref{sec:local-dual-charging}. Fix $t\geq 0$ and $r>0$, and let
\[
  U(r,t):=\{u\in U(t):r\leq t-a_u<2r\}.
\]
As in Section~\ref{sec:local-dual-charging}, a request is old if $a_u<t-3r$ and fresh otherwise.
Deleting the old vertices from the merge tree of an active component
$C$ gives its fresh components. For each fresh component $A$, write
$Q_A:=A\cap U(r,t)$ and $m_A:=|Q_A|$. We only need to charge
components with $m_A>0$. One new issue is that requests in $Q_A$
need not be pairwise separated. We instead use the following $k$-way analog
of Lemma~\ref{lem:separation-of-unmatched-requests}.

\begin{lemma}
\label{app:kway:separation}
If $G$ consists of $k$ unmatched requests in the same active component
at time $t$, then
\[
  \Delta(G)>\frac12\max_{u\in G}(t-a_u).
\]
\end{lemma}
\begin{proof}
Otherwise, $\tau(G)=\min_{u\in G}a_u+2\Delta(G)\leq t$.
The group would be eligible, contradicting the exhaustion of all
eligible groups at time $t$.
\end{proof}
Lemma \ref{app:kway:separation} tells us that the diameter of any $k$ requests in $Q_A$ is greater than $r/2$. We can use this fact to find a large subset of $Q_A$ such that every pair of requests in this subset is separated. 
\begin{lemma}
\label{app:kway:packing}
For every fresh component $A$, there is a set $R_A\subseteq Q_A$ with
$|R_A|\geq m_A/(k-1)$ such that $\widehat d(u,v)>r/4$ for all distinct
$u,v\in R_A$.
\end{lemma}
\begin{proof}
Take a maximal subset $R_A\subseteq Q_A$ whose requests are pairwise
more than $r/4$ apart. Assign every request of $Q_A$ to a member of
$R_A$ within distance $r/4$, which exists by maximality. If $k$
requests were assigned to the same member, their pairwise distances
would be at most $r/2$, while their ages are at least $r$, contradicting
Lemma~\ref{app:kway:separation}. Therefore, each member receives at most $k-1$
requests, proving the bound.
\end{proof}

The rest of the local charging arguments are the same as those in Section \ref{sec:local-dual-charging} but using this separated set $R_A$ instead of $Q_A$. In particular, we construct the witness tree for set $R_A$ and then apply the path charging argument. For the corner case where the component $C$ has a single fresh component $A$ with fewer than $k$ unmatched vertices, we have the following baseline charging lemma (analogous to Lemma \ref{lem:baseline-payment}).
%We also need a baseline charge for a component with fewer than $k$ requests in the age window. 

For a fixed active component $C$ at time
$t$, write $L_C(s):=\sum_{S\subseteq C}y_S(s)$ for $s\leq t$.

\begin{lemma}
\label{app:kway:baseline}
If $u\in U(r,t)$ belongs to an active component $C$ at time $t$, then
$L_C(t)\geq r/64$.
\end{lemma}
\begin{proof}
If the component containing $u$ is growing throughout $[a_u,t]$,
except possibly at event times, then
$L_C(t)\geq\alpha(t-a_u)\geq r/64$.

Otherwise, let $\sigma\leq t$ be the right endpoint of the last
positive-length interval on which the component containing $u$ is
frozen, and let $B$ be that component just before $\sigma$. The
modulo-$k$ invariant implies that $B$ has at least $k$ unmatched
requests, including $u$, on a sufficiently short interval immediately
before $\sigma$. Choose a fixed set $G$ of $k$ such requests containing
$u$. Applying Lemma~\ref{app:kway:separation} before $\sigma$ and
taking a left limit gives $\Delta(G)\geq(\sigma-a_u)/2$.
For a pair $v,w\in G$ attaining $\Delta(G)$, the path between $v$
and $w$ in the merge tree of $B$ has length at least $\widehat d(v,w)$.
The path-dual bound of Lemma~4.15, whose proof is independent of the
growth criterion, therefore gives
\[
  L_C(\sigma)\geq\frac12\widehat d(v,w)\geq\frac{\sigma-a_u}{4}.
\]
After $\sigma$, the component containing $u$ grows at rate $\alpha$
except at event times. All these components are subsets of $C$, so
\[
  L_C(t)\geq\frac{\sigma-a_u}{4}+\alpha(t-\sigma)
    \geq\alpha(t-a_u)\geq\frac{r}{64}.
\]
This argument also applies when $\sigma=t$.
\end{proof}
Finally, putting all these together, we obtain the local dual charging theorem.
\begin{theorem}
\label{app:kway:local-charging}
For every $t\geq 0$ and $r>0$,
\[
  r|U(r,t)|\leq
    64(k-1)\bigl(\Dual(t)-\Dual(t-3r)\bigr).
\]
\end{theorem}
\begin{proof}
Fix an active component $C$ with $C\cap U(r,t)\neq\emptyset$.
If $C$ has no old vertex and $1\leq |C\cap U(r,t)|<k$, its only
fresh component is $A=C$. Lemma~\ref{app:kway:baseline} gives
\[
  L_C(t)\geq\frac{r}{64}\geq\frac{m_A r}{64(k-1)}.
\]
Every dual variable supported on a subset of $C$ was zero at time
$t-3r$, because all requests of $C$ are fresh. This proves the
required recent charge in this case.

In all other cases, choose $R_A$ as in
Lemma~\ref{app:kway:packing} for each fresh component with $m_A>0$.
Root the merge tree of $C$ at an old vertex if one exists, and otherwise
at any vertex. Let $v_A$ be the vertex of $A$ closest to the root.
If $C$ has an old vertex, the parent of $v_A$ is old. The merge-edge
age inequality of Lemma~4.13 is valid for $\alpha=1/64$ and implies
\[
  t-a_{v_A}>3r\frac{1-\alpha}{1+\alpha}
    =\frac{189}{65}r>\frac52r.
\]
Every member of $R_A$ has age less than $2r$, and hence its distance
from $v_A$ is greater than $r/2$, in particular greater than $r/4$.
If $C$ has no old vertex, then $A=C$ and $m_A\geq k$, so
$|R_A|\geq 2$.

Let $W_A$ be the minimal subtree of the merge tree spanning
$R_A\cup\{v_A\}$. It is contained in $A$. Applying the separated
terminal bound of Lemma~4.12 to $R_A\cup\{v_A\}$ in the first
case and to $R_A$ in the second gives
\begin{equation}
  \operatorname{len}(W_A)>\frac{|R_A|r}{8}.
  \label{app:kway:witness-length}
\end{equation}

For completeness, we give the path charging argument with its
dependence on $\alpha$. Root $W_A$ at $v_A$. Successively prune a
path from a non-root leaf to the first branching vertex or the root,
deleting its edges and retaining its other endpoint. Continue along
vertices of degree two in the remaining tree. This decomposes $W_A$
into edge-disjoint paths $P_1,\ldots,P_q$, with $q\leq |R_A|$.
Each path starts at a non-root leaf of the original witness tree,
which is in $R_A$, and ends at an anchor $b_j$ in the remaining tree.
To see that no new non-root leaf is needed, observe that a branching
anchor has degree at least three before a path is deleted and hence
degree at least two afterwards.

For each path, define
\[
  \mathcal D_j:=\{S\in\mathcal H(t):y_S(t)>0,
                   \ S\text{ crosses }P_j,\ b_j\notin S\},
  \qquad
  \mathcal D_A:=\bigcup_{j=1}^q\mathcal D_j,
\]
where $\mathcal H(t)$ is the family of historical components and a
set crosses a path if it separates the endpoints of one of its edges.
Every historical component is connected in the merge tree. Thus a
set charged to a pruned path, but not containing its anchor, cannot
meet any later path. The families $\mathcal D_j$ are consequently
pairwise disjoint. The same connectivity argument shows that no set
in $\mathcal D_A$ contains $v_A$.

The path-dual bound of Lemma~4.15 gives
\[
  \sum_{S\in\mathcal D_j}y_S(t)
    \geq\frac12\operatorname{len}(P_j)
           -\reach(b_j,t).
\]
If $u_j\in R_A$ is the other endpoint of $P_j$, then
$t-a_{b_j}\leq t-a_{u_j}+|a_{u_j}-a_{b_j}|
\leq 2r+\operatorname{len}(P_j)$. By
\eqref{app:kway:reach}, it follows that
\[
  \sum_{S\in\mathcal D_A}y_S(t)
    \geq\left(\frac12-\alpha\right)\operatorname{len}(W_A)
           -2\alpha q r
    >\frac{15}{512}|R_A|r
    \geq\frac{m_A r}{64(k-1)}.
\]
Here we used \eqref{app:kway:witness-length}, $q\leq |R_A|$,
$\alpha=1/64$, and $|R_A|\geq m_A/(k-1)$.

The charge families of distinct fresh components are also disjoint.
Indeed, a connected historical component meeting both $A$ and $A'$
must contain at least one of $v_A$ and $v_{A'}$, whereas membership
in both charge families excludes both vertices. Moreover, every
charged set contains a fresh vertex, so its dual variable was zero
at time $t-3r$. Charge families belonging to different active
components are disjoint as well. Summing all the charges, including
the baseline cases, proves the theorem.
\end{proof}

\subsubsection{Local to global}
\label{app:kway:global}

The preceding bound controls the delay of the algorithm independently
of the group connection objective. We record it separately for use
with minimum spanning tree costs.

\begin{lemma}
\label{app:kway:delay-bound}
The algorithm in Section~\ref{app:kway:algorithm} satisfies
\[
  \Delay\leq
    \frac{64(k-1)}{\ln 2}(1+3\ln n)\Dual.
\]
\end{lemma}
\begin{proof}
Write $B:=64(k-1)$ and $D:=\Dual$. Applying
Theorem~\ref{app:kway:local-charging} with $r=t-a_u$ to a request
that is still unmatched at time $t$ gives $t-a_u\leq BD$.
Taking limits before matching shows that every request waits at most
$BD$. If $D=0$, there is no positive delay and the claim follows.
Assume $D>0$.
We use the same integral identity as in the linear local-to-global
argument of Section~\ref{sec:linear-analysis}:
\[
  \Delay
    =\frac{1}{\ln 2}\int_0^\infty\frac{1}{r}
                         \int_0^\infty |U(r,t)|\,dt\,dr.
\]
For a fixed request, the times at which its age is in $[r,2r)$
form an interval of length at most $r$. Hence
$\int_0^\infty |U(r,t)|\,dt\leq nr$. Also by Lemma \ref{lemma:window-integral} for every $r>0$,
\[
  \int_0^\infty
    \bigl(\Dual(t)-\Dual(t-3r)\bigr)\,dt
       =3rD.
\]
 Therefore, for
$0<\rho\leq BD$, Theorem~\ref{app:kway:local-charging} yields
\begin{align*}
  \Delay
    &\leq\frac{n\rho}{\ln 2}
       +\frac{B}{\ln 2}\int_\rho^{BD}\frac{1}{r^2}
           \int_0^\infty
           \bigl(\Dual(t)-\Dual(t-3r)\bigr)
           \,dt\,dr
    =\frac{n\rho}{\ln 2}+\frac{3BD}{\ln 2}
                         \ln\frac{BD}{\rho}.
\end{align*}
Setting $\rho=BD/n$ proves the lemma.
\end{proof}

\begin{theorem}[Sum costs]
\label{app:kway:sum-theorem}
The Delayed Best-Effort Matching Algorithm of
Section~\ref{app:kway:algorithm} is deterministic and
$O(k\log n)$-competitive for $k$-way matching with delays and
all-pairs sum costs $\operatorname{sum}(G)$. More precisely,
\[
  \ALG\leq
    \frac{96(k-1)}{\ln 2}(1+3\ln n)\OPT.
\]
\end{theorem}
\begin{proof}
Combine Lemmas~\ref{app:kway:sum-dual},
\ref{app:kway:structure-cost}, and \ref{app:kway:delay-bound}.
\end{proof}

\subsection{Other group costs}
\label{app:kway:other-costs}

We first give a comparison principle for group costs that can be
bounded above and below by sum costs.

\begin{proposition}[Approximate-sum group costs]
\label{app:kway:comparison}
Suppose a group connection cost $c(G)$ satisfies
\[
  \lambda\,\operatorname{sum}(G)
       \leq c(G)\leq
  \Lambda\,\operatorname{sum}(G)
\]
for every group of size $k$, where $0<\lambda\leq\Lambda$ may
depend on $k$. Running the algorithm of
Section~\ref{app:kway:algorithm} with spatial metric
$\widetilde d_0=\lambda d_0$ is
$O((\Lambda/\lambda)k\log n)$-competitive for connection cost $c$.
\end{proposition}
\begin{proof}
The sum cost in the scaled metric is
$\widetilde c(G)=\lambda\operatorname{sum}(G)$.
Thus $\widetilde c(G)\leq c(G)\leq
(\Lambda/\lambda)\widetilde c(G)$. Delay costs are unchanged,
and $\Lambda/\lambda\geq1$. Evaluating the output of the scaled
algorithm under either objective therefore gives
$\ALG_c\leq(\Lambda/\lambda)\ALG_{\widetilde c}$,
while $\OPT_{\widetilde c}\leq\OPT_c$.
The result follows from Theorem~\ref{app:kway:sum-theorem}.
\end{proof}

In this application, the algorithm's auxiliary distance is
$\widehat d(u,v)=(k-1)\lambda d_0(x_u,x_v)+|a_u-a_v|$.
The objective being approximated remains the original group cost $c$.

\subsubsection{Diameter costs}
\label{app:kway:diameter-section}

Let $\operatorname{diam}(G):=\max_{u,v\in G}d_0(x_u,x_v)$.

\begin{corollary}
\label{app:kway:diameter}
There is a deterministic $O(k^2\log n)$-competitive algorithm for
$k$-way matching with delays and diameter costs.
\end{corollary}
\begin{proof}
Choose $u^\star,v^\star\in G$ attaining the diameter. For each
$w\in G\setminus\{u^\star,v^\star\}$, the triangle inequality
gives
$d_0(x_{u^\star},x_w)+d_0(x_w,x_{v^\star})
\geq\operatorname{diam}(G)$. Summing these inequalities and the
contribution of $\{u^\star,v^\star\}$ yields
\[
  (k-1)\operatorname{diam}(G)\leq\operatorname{sum}(G)
    \leq\binom{k}{2}\operatorname{diam}(G).
\]
Equivalently,
\[
  \frac{2}{k(k-1)}\operatorname{sum}(G)
    \leq\operatorname{diam}(G)
    \leq\frac{1}{k-1}\operatorname{sum}(G).
\]
Apply Proposition~\ref{app:kway:comparison} with
$\lambda=2/(k(k-1))$ and $\Lambda=1/(k-1)$.
Their ratio is $k/2$. The input spatial metric is
$\widetilde d_0=2d_0/(k(k-1))$, so the algorithm uses auxiliary
distance $\widehat d(u,v)=(2/k)d_0(x_u,x_v)+|a_u-a_v|$.
\end{proof}

\subsubsection[H-metrics]{$H$-metrics}
\label{app:kway:hmetrics}

For a tuple $\mathbf x\in X^k$, let
$\operatorname{supp}(\mathbf x)$ be the set of its distinct entries.
An $H$-metric is a symmetric, nonnegative function
$H:X^k\to\mathbb R_{\geq0}$ that is zero if and only if all
entries are equal, and satisfies
\begin{equation}
\begin{aligned}
  H(\mathbf x)&\leq H(\mathbf z)
     &&\text{if }\operatorname{supp}(\mathbf x)
                    \subsetneq\operatorname{supp}(\mathbf z),\\
  H(\mathbf x)&\leq(k-1)H(\mathbf z)
     &&\text{if }\operatorname{supp}(\mathbf x)
                    =\operatorname{supp}(\mathbf z).
\end{aligned}
\label{app:kway:h-support}
\end{equation}
It also satisfies the generalized triangle inequality: for any
$x_1,\ldots,x_k,z\in X$ and $1\leq i<k$,
\[
  H(x_1,\ldots,x_k)
    \leq H(x_1,\ldots,x_i,\underbrace{z,\ldots,z}_{k-i})
       +H(\underbrace{z,\ldots,z}_{i},x_{i+1},\ldots,x_k).
\]
For a group $G$, $H(G)$ means $H$ evaluated on its request
locations, with repetitions retained.

\begin{corollary}
\label{app:kway:hmetric}
There is a deterministic $O(k^3\log n)$-competitive algorithm for
$k$-way matching with delays and $H$-metric costs.
\end{corollary}
\begin{proof}
Use the pairwise metric associated with $H$ by Kakimura and
Nakayoshi (2025, Theorem~1):
\[
  d_H(x,y):=
     H(x,\underbrace{y,\ldots,y}_{k-1})
      +H(y,\underbrace{x,\ldots,x}_{k-1}).
\]
%For completeness, symmetry and positive definiteness follow from
%the corresponding properties of $H$. 
%Applying the generalized
%triangle inequality with $i=1$ gives
%\[
%   H(x,\underbrace{y,\ldots,y}_{k-1})
%     \leq H(x,\underbrace{z,\ldots,z}_{k-1})
%           +H(z,\underbrace{y,\ldots,y}_{k-1}).
% \]
% Adding the analogous inequality with $x$ and $y$ exchanged proves
% the triangle inequality for $d_H$.

Let
$S_H(G):=\sum_{\{u,v\}\subseteq G}d_H(x_u,x_v)$.
For each pair of locations in $G$, the support conditions
\eqref{app:kway:h-support} give
$d_H(x_u,x_v)\leq2(k-1)H(G)$; this is also true when the locations
coincide. Consequently,
\begin{equation}
  S_H(G)\leq k(k-1)^2H(G).
  \label{app:kway:h-lower}
\end{equation}
For the opposite bound, fix a request $u\in G$ and apply the
generalized triangle inequality repeatedly with bridge point $x_u$.
This gives
\[
  H(G)\leq\sum_{v\in G\setminus\{u\}}
           H(x_v,\underbrace{x_u,\ldots,x_u}_{k-1}).
\]
Sum over all $k$ choices of $u$. The two ordered terms associated
with each unordered pair add up to $d_H(x_u,x_v)$, so
\begin{equation}
  kH(G)\leq S_H(G).
  \label{app:kway:h-upper}
\end{equation}
The two bounds give a comparison directly with the sum cost in
the metric $d_H$:
\[
  \frac{1}{k(k-1)^2}S_H(G)
     \leq H(G)\leq
  \frac{1}{k}S_H(G).
\]
Proposition~\ref{app:kway:comparison} applies with
$\lambda=1/(k(k-1)^2)$ and $\Lambda=1/k$ and yields the claim. 
% Explicitly, the input spatial metric
% is $d_H/(k(k-1)^2)$, so the algorithm's auxiliary distance is
% \[
%   \widehat d(u,v)=\frac{d_H(x_u,x_v)}{k(k-1)}+|a_u-a_v|.
% \]
\end{proof}

\subsubsection{Minimum spanning tree costs}
\label{app:kway:mst-section}

Let
\[
  \operatorname{MST}(G):=
     \min_{T\text{ spanning tree on }G}
       \sum_{\{u,v\}\in E(T)}d_0(x_u,x_v).
\]
For this objective, we obtain a stronger bound by using a direct
dual certificate. We run the same algorithm with input spatial metric
$d_0/(k-1)$, so that its auxiliary distance is
\[
  \widehat d_{\operatorname{MST}}(u,v)
       =d_0(x_u,x_v)+|a_u-a_v|.
\]
The local charging and delay analysis apply to this execution as well.

\begin{corollary}
\label{app:kway:mst}
The algorithm of Section~\ref{app:kway:algorithm}, run on the
input spatial metric $d_0/(k-1)$, is $O(k\log n)$-competitive for
$k$-way matching with delays and minimum spanning tree costs.
Here the MST cost is measured in the original metric $d_0$.
More precisely,
\[
  \ALG_{\operatorname{MST}}\leq
    \frac{128(k-1)}{\ln 2}(1+3\ln n)
      \OPT_{\operatorname{MST}}.
\]
\end{corollary}
\begin{proof}
Let $y$ and $\Dual$ denote the dual variables and their
total value in this execution. We show that these variables satisfy the group
constraints with right-hand side $\operatorname{MST}(G)+W(G)$.
Fix $G$. Double the edges of a minimum spanning tree and shortcut
an Euler tour to obtain a cyclic ordering $u_1,\ldots,u_k$ of its
requests, with $u_{k+1}=u_1$, such that
\[
  \sum_{i=1}^k d_0(x_{u_i},x_{u_{i+1}})
     \leq2\operatorname{MST}(G).
\]
For $k=2$, the cycle traverses the same pair twice. Every set $S$
with $0<|S\cap G|<k$ separates at least two consecutive pairs of
the cycle, counted with multiplicity. Moreover,
\[
  \sum_{i=1}^k|a_{u_i}-a_{u_{i+1}}|
     \leq\sum_{i=1}^k
       \bigl((a_G^{\max}-a_{u_i})+(a_G^{\max}-a_{u_{i+1}})\bigr)
     =2W(G).
\]
Summing the pairwise constraints along the cycle gives
\[
  2\sum_{S:\,0<|S\cap G|<k}y_S
      \leq\sum_{i=1}^k \widehat d_{\operatorname{MST}}(u_i,u_{i+1})
      \leq2\operatorname{MST}(G)+2W(G).
\]
Thus $y$ is feasible for the group-cut dual with MST costs, and
weak duality yields
$\Dual\leq\OPT_{\operatorname{MST}}$.

Let $\Delta_{\operatorname{MST}}(G)
 :=\max_{u,v\in G}\widehat d_{\operatorname{MST}}(u,v)$.
Each group formed at time $t$ satisfies
\[
  \operatorname{MST}(G)\leq(k-1)\Delta_{\operatorname{MST}}(G)
    \leq\frac{k-1}{k+1}\sum_{u\in G}(t-a_u),
\]
by \eqref{app:kway:group-delay}, applied to this execution. Hence
$\ALG_{\operatorname{MST}}\leq2\Delay$.
Combining this inequality with Lemma~\ref{app:kway:delay-bound}
for the same execution
and the dual lower bound proves the corollary.
\end{proof}

\section{Necessity of slow dual growth}
\label{sec:fast-dual-counterexample}

We first show that an online dual growing at rate $1$ can be a polynomial factor smaller than the hindsight optimum. Then in the following subsection we show that reducing the rate to any constant $<1$ bounds the primal-dual gap by a constant.

\subsection{A lower bound on the primal-dual gap for growth rate $1$}

We construct an instance of $\mpmd$ on a line metric where the online dual that grows active components at rate 1 has a dual objective that is $\Theta(\sqrt{n})$ smaller than the cost of the optimal solution.

Let $k$ be a fixed even integer. We consider the underlying metric space to be the line metric and requests arrive at integer locations on the line. The time-augmented metric space in this case is thus the two dimensional Euclidean plane with the $\ell_1$ (i.e., Manhattan) distance metric. We identify a request that arrives at location $i$ and time $t$ by the two dimensional point $(i, t)$. Let $[k] = \{1, 2, \ldots, k\}$. 

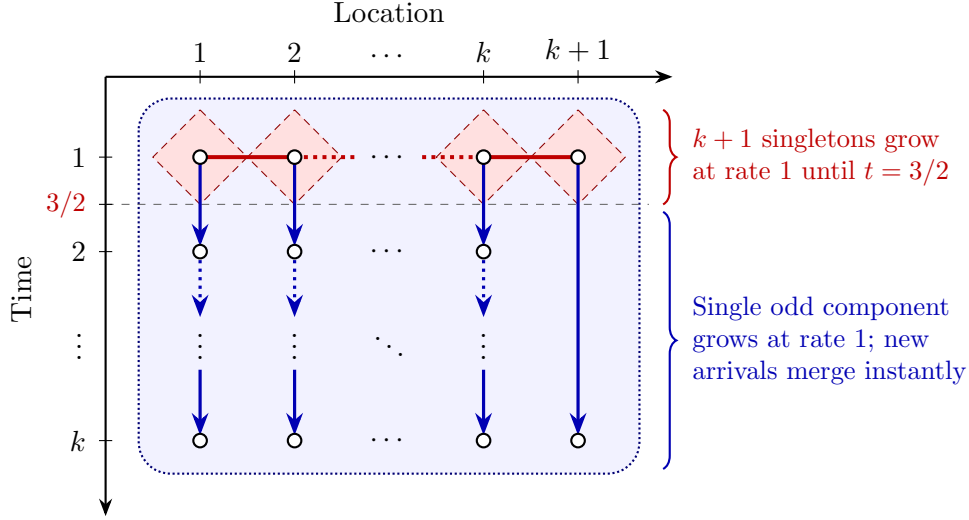
\begin{figure}
    \centering
    \begin{tikzpicture}[
    scale=1.25,
    req/.style={circle, draw=black, fill=white, thick, inner sep=1.8pt},
    mergeedge/.style={draw=red!75!black, line width=1.5pt},
    instantedge/.style={draw=blue!70!black, line width=1.2pt, -Stealth, shorten >=2pt}
]
% Axes
\draw[-Stealth, thick] (0, 0.35) -- (0, -4.3);
\node[rotate=90, anchor=south] at (-0.68, -1.9) {Time};
\draw[-Stealth, thick] (0, 0.35) -- (6.0, 0.35);
\node[anchor=south] at (3.0, 0.82) {Location};
% Location ticks
\foreach \i/\label in {1/1, 2/2, 4/k, 5/k+1} {
    \draw (\i, 0.28) -- (\i, 0.42);
    \node[anchor=south] at (\i, 0.4) {$\label$};
}
\node at (3, 0.56) {$\cdots$};
% Time ticks
\foreach \y/\tlabel in {-0.5/1, -1.5/2, -3.5/k} {
    \draw (-0.07, \y) -- (0.07, \y);
    \node[anchor=east] at (-0.1, \y) {$\tlabel$};
}
\node at (-0.28, -2.42) {$\vdots$};
% t = 3/2 dashed line
\draw[dashed, gray!80!black] (0, -1.0) -- (5.75, -1.0);
\draw (-0.07, -1.0) -- (0.07, -1.0);
\node[anchor=east, font=\small, text=red!75!black] at (-0.1, -1.0) {$3/2$};
% Background shading for L1 moats (diamonds) and the growing odd component
\begin{scope}[on background layer]
    % Large odd component region
    \filldraw[rounded corners=12pt, fill=blue!5, draw=blue!45!black, densely dotted, thick]
        (0.35, 0.12) rectangle (5.65, -3.85);
    % Singleton L1 moats at t=1 (diamonds of L1-radius 1/2)
    \foreach \i in {1, 2, 4, 5} {
        \filldraw[fill=red!12, draw=red!55!black, densely dashed]
            (\i, 0.0) -- (\i+0.5, -0.5) -- (\i, -1.0) -- (\i-0.5, -0.5) -- cycle;
    }
\end{scope}
% Merge edges at t = 3/2 (horizontal along row t=1)
\draw[mergeedge] (1, -0.5) -- (2, -0.5);
\draw[mergeedge, dotted] (2, -0.5) -- (2.65, -0.5);
\draw[mergeedge, dotted] (3.35, -0.5) -- (4, -0.5);
\draw[mergeedge] (4, -0.5) -- (5, -0.5);
% Vertical immediate merge edges at t = 2, ..., k
\foreach \i in {1, 2, 4} {
    \draw[instantedge] (\i, -0.5) -- (\i, -1.5);
    \draw[instantedge, dotted] (\i, -1.5) -- (\i, -2.25);
    \draw[instantedge] (\i, -2.75) -- (\i, -3.5);
}
% Vertical immediate merge edge for (k+1, k) from (k+1, 1)
\draw[instantedge] (5, -0.5) -- (5, -3.5);
% Grid of nodes
% Row t = 1
\foreach \i in {1, 2, 4, 5} {
    \node[req] (v\i1) at (\i, -0.5) {};
}
\node at (3, -0.5) {$\cdots$};
% Row t = 2
\foreach \i in {1, 2, 4} {
    \node[req] (v\i2) at (\i, -1.5) {};
}
\node at (3, -1.5) {$\cdots$};
% Row t = dots
\foreach \i in {1, 2, 4} {
    \node at (\i, -2.42) {$\vdots$};
}
\node at (3, -2.42) {$\ddots$};
% Row t = k
\foreach \i in {1, 2, 4, 5} {
    \node[req] (v\i4) at (\i, -3.5) {};
}
\node at (3, -3.5) {$\cdots$};
% Explanatory braces on the right
\draw[decorate, decoration={brace, amplitude=5pt}, thick, red!75!black]
    (5.9, -0.0) -- (5.9, -1.0)
    node[midway, right=7pt, align=left, font=\small, text=red!75!black]
    {$k+1$ singletons grow\\at rate $1$ until $t = 3/2$};
\draw[decorate, decoration={brace, amplitude=5pt}, thick, blue!70!black]
    (5.9, -1.08) -- (5.9, -3.8)
    node[midway, right=7pt, align=left, font=\small, text=blue!70!black]
    {Single odd component\\grows at rate $1$; new\\arrivals merge instantly};
\end{tikzpicture}
    \caption{Lower-bound instance of $n = k^2 + 2$ requests on the line metric when duals grow at rate $\alpha = 1$. At $t = 1$, the $k+1$ requests grow singleton duals of $1/2$ (shaded red diamonds in the $\ell_1$ metric) until $t = 3/2$, when adjacent horizontal edges (red) become tight and merge them into a single odd active component. Thereafter, this single odd component grows at rate $1$, so all subsequent arrivals at $t = 2, \dots, k$ are immediately absorbed upon arrival via tight vertical edges (blue arrows) without growing singleton duals.}
\label{fig:rate-one-gap}
\end{figure}

The instance then contains the following $(k^2 + 2)$ requests $\{(i, j)\}_{i\in[k], j\in[k]} \cup \{(k+1, 1), (k+1, k)\}$. In other words, first requests arrive at locations $1, \ldots, k+1$ at time $1$. Then at each integer time $t = 2, \ldots, k-1$, requests arrive at the first $k$ locations. Finally at time $k$, requests arrive at locations in $[k+1]$. See Figure \ref{fig:rate-one-gap} for an illustration.

First, note that the instance has exactly $k^2 + 2$ requests and each request is at distance exactly $1$ (in the time augmented metric space) from its nearest neighbor. Thus the optimal solution has cost at least $\frac{k^2 + 2}{2}$.

We now walk through the dual growth process. At time $1$, when the $k+1$ requests arrive, they each form a singleton active component and their corresponding dual variables grow at rate 1. At time $3/2$, the edges between adjacent requests become tight and all the requests merge into a single active component with $k+1$ nodes (odd). So until time $3/2$, the total dual accumulated is exactly $\frac{k+1}{2}$. Immediately after time $3/2$, the single odd active component continues to increase its dual at rate $1$. At time $2$, when a request $(i, 2)$ appears, the edge between $(i, 1)$ and $(i, 2)$ becomes immediately tight. Thus all the $k$ new requests that arrived at time $2$ get merged into the single active component and the dual variables of the singleton active components of these requests never get a chance to increase. This process continues until time $k$, i.e., there remains a single odd active component and
the newly arriving requests merge into this active component as soon as they arrive. Thus from time $3/2$ to time $k$, the total dual only increases at rate $1$ for a total increase of at most $k$. From this point onwards, the active component is even and no further dual growth occurs. The total dual objective is therefore only at most $\Theta(k)$.

Since in this instance we have $n = \Theta(k^2)$, we get $\Dual \leq O(1/\sqrt{n})\ \OPT$.

\subsection{An upper bound on the primal-dual gap for growth rate $\alpha <  1$}

We will prove that if the dual grows at a rate of $\alpha<1$, we have $\OPT \le 2/(1-\alpha)\Dual$.

Consider the merge forest $F=F(\infty)$ of the dual. Note that all of the connected components in this forest have even size. While there exists a connected component $T$ in $F$ and an edge $e\in T$ such that the two connected components in $T\setminus\{e\}$ are both even, drop such an edge, and repeat. The resulting set of edges, $F'$, forms a forest in which every tree is of even size and every vertex has odd degree. We can therefore decompose each component of $F'$ into edge disjoint paths such that every vertex is an end point of exactly one path. Short-cutting these paths provides a perfect matching, and therefore we can conclude that $\OPT\le \len(F')$.

We will now argue that the dual pays a constant fraction of the cost of $F'$. In order to do so, we consider a time $t$, and we want to bound $\sum_{S \text{ active at } t; |S| \text{ odd}} |\delta_{F'}(S)|$. The problem is that $\delta_{F'}(S)$ needs to also account for edges $e$ that have not arrived at time $t$ but are nevertheless loaded by some active $S$ at that time. These are precisely the edges $e=(u,v)$ for which one of the end points arrived before $t$ and the other after.

Let $o(t)$ denote the number of active odd dual components at time $t$. Let $b(t)$ denote the number of vertices that arrive after $t$, but that have at least one neighbor in $F'$ with an arrival time $\le t$.
%and $b(s)$ denote the number of edges of $F'$ that have one endpoint with arrival time $\le s$ and the other with arrival time $>s$. 
We then claim the following lemma.
\begin{lemma}
    At any time $t$, 
    \[\sum_{S \text{ active at } t; |S| \text{ odd}} |\delta_{F'}(S)| \le 2o(t) + b(t)\]
\end{lemma}
\noindent
Before we prove the lemma, we will argue that it leads to a bound on the primal cost. We have:
\begin{align*}
    \len(F') & = \sum_S y_S |\delta_{F'}(S)|
    = \alpha \int \sum_{S \text{ active at } t; |S| \text{ odd}} |\delta_{F'}(S)| dt \\
    & \le \alpha \int (2o(t)+b(t)) dt
    = 2\Dual + \alpha \int b(t)dt
\end{align*}

Noting that $\int b(t)dt \le \sum_{(u,v)\in F'} |a_u-a_v|\le\len(F')$, we get:
\[\OPT\le\len(F')\le \frac{2}{1-\alpha}\Dual.\]

\noindent
It remains to prove the lemma.
\begin{proof}
Let $\mathcal{O}(t)$ and $\mathcal{E}(t)$ denote the set of odd active components and even active components at time $t$ respectively. Remove from $\mathcal{E}(t)$ all components $S$ with $\delta_{F'}(S)=0$, and contract each component in $\mathcal{E}(t)$ and $\mathcal{O}(t)$ into a single meta node. Since each historical dual component is connected in $F$, over the contracted meta nodes, $F$ still forms a forest. Since $F'$ is a subset of $F$, $F'$ is also a forest over the meta nodes. 

Let $F'_t$ denote the subforest of $F'$ containing all edges that have at least one endpoint with arrival time $\le t$. Note that every edge of $F'$ with an endpoint in $\mathcal{O}(t)$ or $\mathcal{E}(t)$ belongs to $F'_t$. Consider the set of all edges in $F'_t$ that have one endpoint with an arrival time before $t$ and the other after $t$; let $\mathcal{B}(t)$ denote the set of endpoints that have not yet arrived. Then $\mathcal{O}(t)$, $\mathcal{E}(t)$, and $\mathcal{B}(t)$ partition the endpoints of all of the edges in $F'_t$.

Observe that the degree of every vertex in $F'$ is odd by construction, and therefore, the degree of all odd sets $\mathcal{O}(t)$ in $F'_t$ is odd, the degree of all even sets $\mathcal{E}(t)$ in $F'_t$ is even (in particular, at least $2$), and the degree of all vertices in $\mathcal{B}(t)$ is at least $1$. Moreover, the average degree of the meta nodes in this forest is at most $2$.

Therefore, using $o(t)$, $e(t)$, and $b(t)$ to denote the sizes of $\mathcal{O}(t)$, $\mathcal{E}(t)$, and $\mathcal{B}(t)$, respectively, we get,
\begin{align*}
\sum_{S\in \mathcal{O}(t)\cup\mathcal{E}(t)\cup\mathcal{B}(t)} |\delta_{F'_t}(S)| & \le 2o(t) + 2e(t) + 2b(t) \\
\implies \sum_{S\in \mathcal{O}(t)} |\delta_{F'}(S)| = \sum_{S\in \mathcal{O}(t)} |\delta_{F'_t}(S)| & \le 2o(t) + 2e(t) + 2b(t) - \sum_{S\in \mathcal{E}(t)} |\delta_{F'_t}(S)|
-\sum_{S\in \mathcal{B}(t)} |\delta_{F'_t}(S)|\\
& \leq 2o(t) + 2e(t) + 2b(t) - 2e(t) - b(t) = 2o(t) + b(t)
\end{align*}
where we used the fact that $|\delta_{F'_t}(S)|\ge 2$ for all $S\in\mathcal{E}(t)$ and $|\delta_{F'_t}(S)|\ge 1$ for all $S\in\mathcal{B}(t)$.
\end{proof}
\end{document}